\documentclass[iop, 
letter]{aastex62}
\usepackage{amsmath}     
\usepackage{wasysym}           
\usepackage{graphicx}
\usepackage{amssymb}
\usepackage{epstopdf}
\usepackage{mathrsfs}
\usepackage{anyfontsize}
\usepackage{natbib}
\usepackage{color}
\usepackage{lipsum}
\usepackage{diagbox}

\shorttitle{Self-similar shock propagation}
\shortauthors{Coughlin et. al}
\begin{document}
\title{Weak Shock Propagation with Accretion.~I.~Self-similar Solutions and Application to Failed Supernovae}
\author[0000-0003-3765-6401]{Eric R. Coughlin}
\altaffiliation{Einstein Fellow}
\author{Eliot Quataert}
\affiliation{Astronomy Department and Theoretical Astrophysics Center, University of California, Berkeley, Berkeley, CA 94720}
\author{Stephen Ro}
\affiliation{Astronomy Department and Theoretical Astrophysics Center, University of California, Berkeley, Berkeley, CA 94720}

\email{eric\_coughlin@berkeley.edu}

\begin{abstract}
We present solutions for the self-similar propagation of a shock wave in a hydrostatic, adiabatic medium with a point mass gravitational field. In contrast to the well-known, Sedov-Taylor blastwave, these solutions apply to the case when the shock Mach number is of order a few, and the energy of the shocked fluid is not conserved but self-consistently modified by the binding energy of the ambient medium that is swept up by the passage of the shock. Furthermore, we show that there is one solution (for a given ambient density profile) that smoothly passes through a sonic point in the post-shock flow and results in accretion onto the central object; in analogy with the Bondi problem, we propose that these solutions are the ones that are most relevant in astrophysical environments. We apply these accreting models to failed supernovae, in which neutron star formation does not unbind the envelope, but a weak shock is still generated in the outer layers of the star from neutrino-induced mass loss. We find excellent agreement between the predictions of our self-similar, shock propagation model and numerical simulations of the collapse of a yellow supergiant; the self-similar solutions reproduce the overall scaling of the shock speed, the time and space-dependent evolution of the velocity, density, and pressure behind the shock, and the accretion rate onto the black hole. Our results have important implications for the fallback and ejection of material in failed supernovae.
\end{abstract}


\section{Introduction}
\label{sec:introduction}
The generation and propagation of a shock wave is key for understanding successful, core-collapse supernovae: in the limit that the energy associated with the neutron star bounce far outweighs the gravitational energy of the star, the shock generated from the bounce propagates through and unbinds the stellar envelope. 
 When the density profile of the progenitor can be approximated as a power-law, the well-known, self-similar, Sedov-Taylor blastwave \citep{sedov59, taylor50} describes the propagation of the shock and the temporal evolution of the post-shock fluid; as long as the density of the ambient medium does not fall off steeper than $\rho \propto r^{-3}$, the shock decelerates during the blastwave phase.\footnote{When the density falls off steeper than $\rho \propto r^{-3}$, a different phase of self-similar propagation is realized owing to the divergence of the energy of the Sedov-Taylor solution and the existence of a critical point within the flow; see \citet{koo90} and \citet{waxman93}.} As the shock approaches the stellar surface, a new regime of self-similar shock propagation can occur if the density falls off approximately as $\rho \propto (1-r/R_*)^{n}$, with $n$ a positive number and $R_*$ the stellar radius; as shown by \citet{sakurai60}, the self-similar solution predicts the acceleration of the shock as it runs down the density gradient of the envelope, and the precise nature of the acceleration depends on the adiabatic indices of the pre and post-shock gas and the power-law index $n$ \citep{ro13}. In between these limits of acceleration and deceleration, there is no known similarity solution; nevertheless, \citet{matzner99} demonstrated that there is a simple, analytic prescription that interpolates between the two self-similar states and agrees very well with the shock evolution found in simulations of core-collapse supernovae.
 
There are two fundamental assumptions underlying the Sedov-Taylor and Sakurai similarity solutions: a strong shock, so that the Mach number is large enough that the pressure of the ambient medium can be ignored in the jump conditions, and the neglect of any gravitational fields.\footnote{The Sakurai solution also adopts a Cartesian geometry, and thus deals only with the very edge of the star.} There are, however, astrophysically-relevant scenarios where these assumptions are not justified. 

One such a scenario is a \emph{failed supernova} -- when the shock generated from neutron star formation is insufficient to overcome the ram pressure of the infalling envelope and the energy lost to nuclear dissociation, and the continued accretion onto the core leads directly to the formation of a black hole. Even when the bounce shock is stifled, however, the star responds and expands dynamically owing to the large amount of mass-energy radiated in the form of neutrinos during protoneutron star formation, and a sound pulse is launched from small radii that steepens into a shock in the outer regions of the star \citep{nadezhin80, lovegrove13, piro13, lovegrove17, fernandez18, coughlin18}. When the stellar progenitor is a supergiant, the shock generated from the neutrino mass loss accelerates only briefly at the edge of the helium core, reaching a Mach number of only a few, before encountering the extended, rarefied hydrogen envelope of the star. In this case, because the hydrogen envelope is in hydrostatic equilibrium, the modest value of the Mach number implies not only that the strong shock assumption is invalid, but that the gravitational potential energy of the fluid is not always negligible in comparison to the shock kinetic energy. Thus, in this regime of shock propagation, it is likely that neither the Sedov-Taylor nor the Sakurai similarity solution is realized. 

In this paper, we show that there exists another class of self-similar solution that describes the propagation of this low to intermediate-Mach number shock in a gravitational field. In particular, if the ambient medium is hydrostatic, adiabatic, and its gravitational field is dominated by the mass of a central object -- very close to the conditions in the hydrogen envelope of a supergiant -- then the shock jump conditions and the fluid equations can be satisfied by self-similar scalings for the velocity, density, and pressure of the post-shock fluid. Interestingly, we find that these self-similar solutions admit not only outward motion immediately behind the shock, but infall and \emph{accretion} onto the central object after a finite time. Furthermore, we demonstrate that there is only \emph{one} value of the shock velocity that satisfies both the jump conditions and permits the smooth passage of the fluid quantities through a sonic point within the flow, meaning that the self-similar solution is uniquely specified by the properties of the ambient medium and the mass of the accreting object. {}{We plan to investigate, in a series of followup papers, the stability of these solutions and how they transition to the Sedov-Taylor regime.}

In Section \ref{sec:scalings} we give the basic scalings for the physical quantities of the fluid (e.g., shock velocity, Mach number, energy) expected to be upheld when the self-similarity of the flow is maintained. In Section \ref{sec:equations} we write down the relevant equations of motion and present the self-similar solutions that match onto the shock jump conditions. Section \ref{sec:simulations} compares the predictions of our self-similar model to the results of numerical simulations of the propagation of a shock -- originally generated from a failed supernova -- through the hydrogen envelope of yellow and red supergiants (YSGs and RSGs), specifically the simulations analyzed in \citet{fernandez18}. We conclude and discuss the implications of our findings in Section \ref{sec:conclusions}.

\section{Basic scalings and assumptions}
\label{sec:scalings}
The Sedov-Taylor blastwave proposes that a large amount of energy, $E$, is injected impulsively into an ambient medium with density profile $\rho \propto r^{-n}$ (for simplicity and because it is relevant for our purposes, we will restrict ourselves to the impulsive case and not allow the energy injection to vary in time). The ``largeness'' of the energy implies that the shock resulting from the energy injection is strong, with Mach number much greater than one, and the energy density of the ambient medium is negligible; the gravitational field -- of a point mass at the origin or of the ambient gas itself -- is also ignored. With these assumptions and if $n < 3$ (see the above footnote about the second-type similarity solutions that apply when $n \ge 3$), there is only one possible radial scaling for the velocity of the shock generated by the explosion, which results from energy conservation and is given by

\begin{equation}
v_{\rm ST}(t) = V\sqrt{\frac{E}{4\pi \rho_0 r_0^3}}\left(\frac{r_{\rm ST}(t)}{r_0}\right)^{\frac{1}{2}\left(n-3\right)} \propto \left(\rho(r_{\rm ST})\,r_{\rm ST}^3\right)^{-1/2}  \label{stvel}
\end{equation}
\begin{equation*}
\Rightarrow \quad r_{\rm ST}(t) \propto t^{\frac{2}{5-n}},
\end{equation*}
where $V$ is a constant and $r_0$ is determined by the ambient medium (and can be renormalized to 1 through a suitable redefinition of $\rho_0$). This simple scaling for the shock velocity suggests that the velocity, density, and pressure behind the shock can be described self-similarly, with the self-similar variable $\xi = r/r_{\rm ST}(t)$, and with boundary conditions for the post-shock flow that are set by the jump conditions at the shock front. The constant $V$ can then be determined by evaluating the energy integral directly in terms of the self-similar functions, and the shock velocity is uniquely specified by the energy of the explosion and the ambient density profile. 

Because the post-shock pressure is much greater than the ambient pressure when the shock is strong, the ambient sound speed does not appear in the Sedov-Taylor solution, and the Mach number of the shock can have any radial scaling. However, if we impose that the ambient medium is in hydrostatic equilibrium prior to the arrival of the shock and that the gravitational field is dominated by mass at small radii (i.e., radii much smaller than the shock location at some initial time), which are reasonable assumptions in many supernova explosions, then the ambient sound speed satisfies

\begin{equation}
c_1^2 \propto \frac{p_1}{\rho_1} \propto \frac{1}{r},
\end{equation}
where subscript 1's refer to the pre-shock gas. In this case, the Mach number of the shock describing the Sedov-Taylor blastwave is

\begin{equation}
\mathscr{M}_{\rm ST} \propto r^{\frac{1}{2}\left(n-2\right)}. \label{stmach}
\end{equation}
Thus, as long as the density of the ambient medium falls off more rapidly than $\rho \propto 1/r^2$, the Mach number of the energy-conserving explosion grows with radius, and the strong shock assumption becomes increasingly justified at late times.

The basic question with which we are concerned here is: is there a separate class of self-similar solution when the energy of the initial blast is comparable to the gravitational energy of the ambient medium? Specifically, when $E/E_{\rm grav} \simeq 1$, where

\begin{equation}
E_{\rm grav} \simeq \int_0^{r}\frac{GM(r)}{r}\rho r^2 dr \propto r^{2-n} \label{egrav}
\end{equation}
is the gravitational binding energy of the material through which the shock is passing (and we assumed that the mass interior to $r$ is roughly constant to obtain the last proportionality), can we still describe the propagation of the shock self-similarly, but with a self-similar variable and a shock velocity that differ from those in the Sedov-Taylor solution?

To motivate the answer to this question, first note that when the ambient energy density is not ignored, then the total energy contained within the blast wave will be time dependent because, as the shock moves out, it will sweep up not just mass -- a familiar feature of the Sedov-Taylor blastwave -- but also energy. There will therefore be a source term in the energy equation for the shocked fluid, proportional to $E_{\rm grav} v_{\rm sh}$, that accounts for the energy flux into the post-shock gas. Since the pre-shock material in any realistic astrophysical environment is bound, we expect that this addition of energy from the ambient medium will reduce the energy content within the blastwave.

Furthermore, if the solution is self-similar, then the object interior to the flow that provides the gravitational field cannot have a finite radius, as otherwise there would be an additional length scale in the problem. This point mass at the center of the flow can also accrete material and remove energy from the fluid within the blast wave, and, for the specific problem of a failed supernova where a black hole is left behind, we \emph{expect} accretion at the center. If the matter is bound at the time of accretion, then the point mass removes negative energy from the shocked gas, providing a source of energy to the overall energy budget of the fluid. {}{}

For these two reasons -- the sweeping up of ambient binding energy and the accretion of energy by the central mass -- we do not expect the energy of the shocked fluid in a weak, self-similar shockwave to be conserved, and the scaling of the velocity will, in general, differ from that of the Sedov-Taylor blastwave. To determine what this scaling of the velocity is, note that when the ambient medium is in hydrostatic equilibrium, the sound speed scales identically to the local free-fall speed. Again, when the self-gravity of the gas can be ignored, this velocity is just a power-law with radius. Additionally, if the shock velocity also scales as the local infall speed, then the internal energy, kinetic energy, and gravitational energy of the post-shock gas all scale identically, and the total energy content of the blastwave evolves as a power-law in time (or radius). There is thus no preferred length scale (or time scale) at which a specific component of the energy will dominate over the others, which would otherwise violate self-similarity. We therefore suggest that, in this case when the energy content of the blast is modified by the ambient medium and accretion, a reasonable scaling for the shock velocity that preserves the self-similarity of the flow is

\begin{equation}
v_{\rm SS} = V\sqrt{\frac{GM}{r_{\rm SS}}} \quad \Rightarrow \quad r_{\rm SS} \propto t^{2/3}, \label{Vss}
\end{equation}
where, as is the case for the Sedov-Taylor solution, $V$ is an undetermined parameter, and the enclosed mass was assumed to be constant to derive the shock position. 

For the Sedov-Taylor solution, the (in principle) arbitrary scaling of the ambient sound speed means that the boundary conditions at the shock front can only be satisfied self-similarly (i.e., such that the post-shock fluid quantities are just constant multiples of the pre-shock quantities) in the strong shock limit. Here, however, this is not the case, because the shock speed and the ambient sound speed both scale as the escape speed, and the shock jump conditions are satisfied in a self-similar manner for an arbitrary shock strength; we demonstrate this feature explicitly in Section \ref{sec:equations}. We stress the importance of this aspect of these solutions: in order for the gravitational energy to be comparable to the kinetic energy, the shock Mach number cannot be much greater than one, and therefore imposing the strong shock limit would not yield meaningful solutions to this problem. Moreover, the identical scaling between the shock velocity and the ambient sound speed implies that the Mach number of the shock is constant:

\begin{equation}
\mathscr{M}_{\rm SS} = const.
\end{equation}
The constancy of the Mach number is a self-consistent feature of these solutions -- a declining Mach number would imply that the gravitational energy eventually far outweighs the kinetic energy, while a growing Mach number would mean that the Sedov-Taylor solution, for which the kinetic energy dominates, eventually takes over at large radii. 

The energy of the shocked fluid for this self-similar solution scales as

\begin{equation}
E_{\rm SS} \propto \rho r^3 v^2 \propto r^{2-n} \propto t^{\frac{4}{3}-\frac{2n}{3}}. \label{Eapp}
\end{equation}
When $n = 2$, the energy is conserved, identical to the Sedov-Taylor solution. However, the case of $n = 2$ is problematic: investigating Equation \eqref{egrav} more carefully shows that the gravitational potential energy of the ambient medium actually diverges logarithmically with this value of $n$. Therefore, in order to successfully propagate to infinity, the shock would need an infinite amount of kinetic energy in order to overcome the infinite binding energy of the ambient medium. This implies that, in the absence of any additional source of energy, the only physical solution when $n = 2$ is the Sedov-Taylor blastwave. Investigating Equation \eqref{egrav} shows that the gravitational potential energy of the ambient gas diverges as a power-law when $n < 2$. Thus, self-similar solutions that satisfy Equations \eqref{Vss} -- \eqref{Eapp} are only possible when the power-law index of the ambient medium satisfies $n \ge2$, with equality reducing to the Sedov-Taylor blastwave.\footnote{The one exception is when the adiabatic index of the ambient medium is $\gamma = 4/3$, for which the ambient energy density is exactly zero {}{owing to the precise balance between the gravitational potential energy and the thermal energy}. We will return to this case briefly in Section \ref{sec:winds}.}

Unlike the Sedov-Taylor blastwave, which uses the energy to infer the shock speed, these self-similar solutions \emph{require} a specific scaling of the energy for a given, dimensionless shock speed $V$. We will also show in Section \ref{sec:equations} that there is a single value of $V$ for a given ambient density profile that permits the smooth passage from the shock front to the origin and results in supersonic accretion. Thus, these self-similar solutions necessitate a single value of the integrated, post-shock energy for a given $n$. 

However, this specific value of the energy may not always be attainable: note from Equation \eqref{egrav} that, when $n > 2$, there is a finite amount of binding energy contained in the ambient medium from any $r_0 > 0$ to infinity. Thus, if we imagine that a shock at radius $r_0$ initially possesses an energy $E$ that satisfies $E - E_{\rm grav} \gtrsim E_{\rm SS}$, then the energy budget of the shockwave cannot ever be brought low enough to equal the self-similar value. The shock will instead transition asymptotically to the Sedov-Taylor solution, with a new energy $\simeq E-E_{\rm grav}$. When this situation occurs in a failed supernova, the generated shock will propagate through the hydrogen envelope with this energy, and upon reaching the surface of the star accelerate in a Sakurai-like phase and unbind the material. Additionally, the initial energy of the shock cannot be too low, as the \emph{most} that can happen is that all of the material bound to the black hole is accreted. If the remaining, positive energy is still below the value required by the self-similar solution, then the gravitational potential energy will eventually dominate the kinetic energy. In this case the Mach number must decline, and eventually the shock dissipates owing to an insufficient energy supply. In a failed supernova, this scenario would result in no material ejected, and the entire envelope would be accreted by the black hole.

When $E - E_{\rm grav} \lesssim E_{\rm SS}$, the larger initial energy correlates with a larger initial shock velocity, which causes ambient energy to be swept up more quickly, reducing the energy budget, and allowing the solution to transition to the self-similar one. If the initial energy is only slightly below the self-similar value, then the smaller shock velocity implies that the binding energy of the ambient medium is swept up at a lower rate, augmenting the total energy behind the shock to the point where the self-similar solution can be achieved. Thus, when the energy is neither too high nor too low, it is likely that the blastwave can self-regulate to the point where the self-similar solution is realized. 

Since the shock velocity scales as the local escape speed, it follows that{}{, if the density varies self-similarly,} the accretion rate onto the central object is

\begin{equation}
\dot{M}_{\bullet} \propto r^2\rho v \propto r^{\frac{3}{2}-n} \propto t^{1-\frac{2n}{3}}. 
\end{equation}
This scaling is identical to the one expected if the gas were in pure freefall, which is reasonable given the nature of the shock velocity. However, because the shock imparts some energy to the gas before it is accreted, the overall normalization of the accretion rate will differ.

To summarize, we expect that a shock with energy comparable to the binding energy of the ambient medium, which has a density profile $\rho \propto r^{-n}$ and is situated in the gravitational field of a central compact object, can be described self-similarly if it has a shock velocity $v_{\rm sh} \propto r^{-1/2}$, a constant Mach number, a total energy $E \propto r^{4/3-2n/3}$, and an accretion rate onto the central mass of $\dot{M}_{\bullet} \propto t^{1-2n/3}$. We further propose that such self-similar propagation is only possible if the power-law index of the ambient medium satisfies $n \ge 2$, as otherwise the shock would need infinite energy. In the following section, we show that there are solutions to the fluid equations with these properties. 

\section{Equations and shock solutions}
\label{sec:equations}
Similar to the Sedov-Taylor blastwave solution, we assume here that a shock is propagating radially outward in a stratified medium and we seek to determine the evolution of the post-shock fluid, which has radial velocity $v$, density $\rho$, and pressure $p$. However, here we approximately account for the gravitational potential energy by assuming that the gas is situated in the gravitational field of a central object, of mass $M$, that dominates the mass of the fluid itself. Adopting spherical symmetry, the continuity, radial momentum, and entropy equations are then

\begin{equation}
\frac{\partial\rho}{\partial t}+\frac{1}{r^2}\frac{\partial}{\partial r}\left(r^2\rho v\right) = 0, \label{cont}
\end{equation}
\begin{equation}
\frac{\partial v}{\partial t}+v\frac{\partial v}{\partial r}+\frac{1}{\rho}\frac{\partial p}{\partial r} = -\frac{GM}{r^2}, \label{rmom}
\end{equation}
\begin{equation}
\frac{\partial s}{\partial t}+v\frac{\partial s}{\partial r} = 0, \label{gasen}
\end{equation}
where 

\begin{equation}
s = \frac{p}{\rho^{\gamma_2}}
\end{equation}
is the specific entropy and $\gamma_2$ is the adiabatic index of the gas in the post-shock fluid. {}{Note that Equation \eqref{gasen} only holds in the post-shock fluid -- there is an entropy jump generated by the shock.}

In the comoving frame of the shock, the gas satisfies the shock jump conditions:

\begin{equation}
\rho_1v_{\rm sh} = \rho_2 \tilde{v}_2,
\end{equation}
\begin{equation}
\rho_1v_{\rm sh}^2+p_1 = \rho_2\tilde{v}_2^2+p_2,
\end{equation}
\begin{equation}
\frac{1}{2}v_{\rm sh}^2+\frac{\gamma_1}{\gamma_1-1}\frac{p_1}{\rho_1} = \frac{1}{2}\tilde{v}_2^2+\frac{\gamma_2}{\gamma_2-1}\frac{p_2}{\rho_2},
\end{equation}
where $v_{\rm sh}$ is the velocity of the shock in the lab frame, subscript 1's denote lab frame ambient quantities, subscript 2's denote lab frame post-shock quantities, and $\tilde{v}_2 = v_{\rm sh}-v_2$ is the post-shock velocity in the comoving frame of the shock; note that $\gamma_2$, the adiabatic index of the post-shock gas, does not necessarily equal the adiabatic index of the pre-shock gas, $\gamma_1$. Solving the above three equations for the post-shock quantities gives (see also \citealt{ostriker88})

\begin{equation}
v_2 = \left\{1-\frac{\gamma_2}{\gamma_2+1}\left(1+\frac{p_1}{\rho_1v_{\rm sh}^2}\right)+\frac{\gamma_2-1}{\gamma_2+1}\sqrt{\left(\frac{\gamma_2}{\gamma_2-1}\right)^2\left(1+\frac{p_1}{\rho_1v_{\rm sh}^2}\right)^2-\left(\frac{\gamma_2+1}{\gamma_2-1}\right)\left(1+\frac{2\gamma_1}{\gamma_1-1}\frac{p_1}{\rho_1v_{\rm sh}^2}\right)}\right\}v_{\rm sh}.
\end{equation}
\begin{equation}
\rho_2 = \frac{\gamma_2+1}{\gamma_2-1}\left\{\frac{\gamma_2}{\gamma_2-1}\left(1+\frac{p_1}{\rho_1v_{\rm sh}^2}\right)-\sqrt{\left(\frac{\gamma_2}{\gamma_2-1}\right)^2\left(1+\frac{p_1}{\rho_1v_{\rm sh}^2}\right)^2-\left(\frac{\gamma_2+1}{\gamma_2-1}\right)\left(1+\frac{2\gamma_1}{\gamma_1-1}\frac{p_1}{\rho_1v_{\rm sh}^2}\right)}\right\}^{-1}\rho_1.
\end{equation}
\begin{equation}
p_2 = \frac{1}{\gamma_2+1}\left\{1+\frac{p_1}{\rho_1v_{\rm sh}^2}+\left(\gamma_2-1\right)\sqrt{\left(\frac{\gamma_2}{\gamma_2-1}\right)^2\left(1+\frac{p_1}{\rho_1v_{\rm sh}^2}\right)^2-\left(\frac{\gamma_2+1}{\gamma_2-1}\right)\left(1+\frac{2\gamma_1}{\gamma_1-1}\frac{p_1}{\rho_1v_{\rm sh}^2}\right)}\right\}\rho_1v_{\rm sh}^2
\end{equation}
The sign here was chosen so that the result agrees with the strong shock limit (and the opposite sign gives a negative post-shock velocity). 

We now assume that the ambient medium is in hydrostatic equilibrium and follows an adiabatic equation of state; additionally imposing that the gravitational field of the central object dominates the self-gravity of the fluid and that we are sufficiently far from the surface of the star, we find that the density and pressure fall off according to

\begin{equation}
\rho_1 = \rho_0\left(\frac{r}{r_0}\right)^{-n},
\end{equation}
\begin{equation}
p_1 = \frac{1}{n+1}\frac{GM}{r}\rho_0\left(\frac{r}{r_0}\right)^{-n},
\end{equation}
where $\rho_0$ is the density at $r_0$. We see that the sound speed of the ambient medium is $c_1^2 \propto p_1/\rho_1 \propto GM/r$, which scales identically to the local escape speed. This suggests that a self-similar solution for the evolution of the post-shock gas can be found if the shock velocity itself scales as the local escape speed; we therefore adopt

\begin{equation}
v_{\rm sh} = V\sqrt{\frac{GM}{r_{\rm sh}}} \quad \Leftrightarrow \quad r_{\rm sh} = \left(\frac{3}{2}V\sqrt{GM}t\right)^{2/3},
\end{equation}
with $V$ an as-yet-undetermined number. With this assumption, the shock jump conditions reduce to

\begin{equation}
v_2 = c_{v}\times V\sqrt{\frac{GM}{r}}, \label{vss2}
\end{equation}
\begin{equation}
\rho_2 = c_{\rho}\times\rho_0\left(\frac{r}{r_0}\right)^{-n},
\end{equation}
\begin{equation}
p_2 = c_p\times\frac{GM}{r}\rho_0 V^2\left(\frac{r}{r_0}\right)^{-n}, \label{pss2}
\end{equation}
where

\begin{equation}
c_v = 1-\frac{\gamma_2}{\gamma_2+1}\left(1+\frac{1}{(n+1)V^2}\right)+\frac{\gamma_2-1}{\gamma_2+1}\sqrt{\left(\frac{\gamma_2}{\gamma_2-1}\right)^2\left(1+\frac{1}{(n+1)V^2}\right)^2-\frac{\gamma_2+1}{\gamma_2-1}\left(1+\frac{2\gamma_1}{(n+1)(\gamma_1-1)V^2}\right)}, \label{cv}
\end{equation}
\begin{equation}
c_\rho = \frac{\gamma_2+1}{\gamma_2-1}\left\{\frac{\gamma_2}{\gamma_2-1}\left(1+\frac{1}{(n+1)V^2}\right)-\sqrt{\left(\frac{\gamma_2}{\gamma_2-1}\right)^2\left(1+\frac{1}{(n+1)V^2}\right)^2-\frac{\gamma_2+1}{\gamma_2-1}\left(1+\frac{2\gamma_1}{(n+1)(\gamma_1-1)V^2}\right)}\right\}^{-1},
\end{equation}
\begin{equation}
c_p = \frac{1}{\gamma_2+1}\left\{1+\frac{1}{(n+1)V^2}+\left(\gamma_2-1\right)\sqrt{\left(\frac{\gamma_2}{\gamma_2-1}\right)^2\left(1+\frac{1}{(n+1)V^2}\right)^2-\frac{\gamma_2+1}{\gamma_2-1}\left(1+\frac{2\gamma_1}{(n+1)(\gamma_1-1)V^2}\right)}\right\}. \label{cp}
\end{equation}
Equations \eqref{vss2} -- \eqref{pss2} motivate the following, self-similar scalings for the post-shock flow variables appearing in Equations \eqref{cont} -- \eqref{gasen}:

\begin{equation}
v = V\sqrt{\frac{GM}{r}}f(\xi), \label{vss}
\end{equation}
\begin{equation}
\rho = \rho_0\left(\frac{r}{r_0}\right)^{-n}g(\xi), \label{rhoss}
\end{equation}
\begin{equation}
p = \frac{GM}{r}\rho_0V^2\left(\frac{r}{r_0}\right)^{-n}h(\xi), \label{pss}
\end{equation}
where

\begin{equation}
\xi = \frac{r}{r_{\rm sh}(t)} = \frac{r}{\left(\frac{3}{2}V\sqrt{GM}t\right)^{2/3}}
\end{equation}
and, from the shock jump conditions, 

\begin{equation}
f(1) = c_v, \quad g(1) = c_\rho, \quad h(1) = c_p. \label{bcs}
\end{equation}
Inserting Equations \eqref{vss} -- \eqref{pss} into Equations \eqref{cont} -- \eqref{gasen} gives the following three, ordinary differential equations:

\begin{equation}
-\xi^{5/2}g'+\left(\frac{3}{2}-n\right)fg+\xi\frac{d}{d\xi}\left(fg\right) = 0, \label{sseq1}
\end{equation}
\begin{equation}
-\xi^{5/2}f'+f\left(-\frac{1}{2}f+\xi f'\right)+\frac{1}{g}\left(-(n+1)h+\xi h'\right) = -\frac{1}{V^2}, \label{sseq2}
\end{equation}
\begin{equation}
-\xi^{5/2}\frac{d}{d\xi}\left(\frac{h}{g^{\gamma_2}}\right)+f\xi\frac{d}{d\xi}\left(\frac{h}{g^{\gamma_2}}\right) +\left(n\gamma_2-n-1\right)\frac{fh}{g^{\gamma_2}}= 0. \label{sseq3}
\end{equation}

Equations \eqref{sseq1} -- \eqref{sseq3} are the three, fundamental, ordinary differential equations that govern the self-similar velocity, density, and pressure of the post-shock fluid. {}{We note that our convention in defining the self-similar functions $f$, $g$, and $h$ -- specifically writing them in terms of Eulerian $r$ and the self-similar variable $r / r_{\rm sh}(t)$ -- was motivated by the fact that the gravitational field is itself a function of Eulerian $r$. However, most other investigations of self-similar shock propagation (which are in the absence of gravitational fields) instead work in the variables $t$ and $r/r_{\rm sh}(t)$. For the sake of completeness, in the Appendix we derive the equations one would obtain if the variables $\{t,r/r_{\rm sh}(t)\}$ were used in place of $\{r,r/r_{\rm sh}(t)\}$, which are formally identical to Equations \eqref{sseq1} -- \eqref{sseq3} under a suitable redefinition of the functions $f$, $g$, and $h$.}

{}{Before moving on to analyze the solutions to Equations \eqref{sseq1} -- \eqref{sseq3}, we emphasize that self-similar solutions exist for general combinations of $\{\gamma_1,\gamma_2,n\}$. However, for our intended astrophysical application of shock propagation through the envelope of a supergiant (see Section \ref{sec:simulations}), one expects the physical range of values adopted by these three parameters to be more restricted. For one, we will show in the next subsection that, for most values of $n$, the shock Mach number of the self-similar solutions is not much greater than one. The gas microphysics and the relative importance of radiation pressure compared to gas pressure should therefore be roughly unaltered in going from the pre to post-shock fluid, and hence $\gamma_1 \simeq \gamma_2 \equiv \gamma$. Second, the envelopes of most supergiants are convectively unstable because of their large luminosities and low effective temperatures, and owing to the fact that the efficiency of convective energy transport is usually quite large, the true pressure gradient through the envelope of the star should be very close to the adiabatic one (e.g., \citealt{hansen04}). The entropy gradient throughout the hydrogen envelope should therefore be quite small, and hence $p \simeq \rho^{\gamma}$. Using this relation in the equation of hydrostatic equilibrium then shows $\rho \propto r^{-n}$, where $n = 1/(\gamma-1)$. }

{}{For these reasons, in most of what follows we will significantly reduce the parameter space of our solutions and simplify the notation by letting $\gamma_1 = \gamma_2 = \gamma = 1+1/n$. In Table \ref{table:1}, for ease of reference of the reader, we define these quantities.}

\begin{table}
\begin{center}
\begin{tabular}{|c|c|c|c|}
\hline
& {\small Adiabatic index of ambient gas} & {\small Adiabatic index of shocked gas} & {\small Power-law index of ambient density} \\
\hline
{\small General} & $\gamma_1$ & $\gamma_2$ & $n$ \\ 
\hline
{\small Supergiant envelope} & $\gamma_1 = \gamma = 1+1/n$ & $\gamma_2 = \gamma = 1+1/n$ & $n$ \\
\hline
\end{tabular}
\end{center}
\caption{{}{A summary of the variables used in the self-similar solutions, being the adiabatic index of the ambient medium, $\gamma_1$, the adiabatic index of the shocked gas, $\gamma_2$, and the power-law index of the density of the ambient gas, $n$ (so $\rho \propto r^{-n}$, where $\rho$ is the density of the ambient medium). Solutions exist for general combinations of these three quantities, but when the shock propagates through the envelope of a supergiant, which is our intended application of the self-similar solutions, physical considerations imply that $\gamma_1 \simeq \gamma_2 \simeq 1+1/n$. }}
\label{table:1}
\end{table}

\subsection{Sonic point}
Equations \eqref{sseq1} -- \eqref{sseq3} can be written in the form

\begin{equation}
\mathbf{M}\cdot\frac{d\mathbf{f}}{d\xi} = \mathbf{R},
\end{equation}
where ${}\mathbf{f} = \left(f,g,h\right)^{{}{T}}$, and $\mathbf{M}$ and $\mathbf{R}$ are a matrix and vector that depend on $f$, $g$, $h$, and $\xi$. It can be shown that

\begin{equation}
\det(\mathbf{M}) \propto \frac{\gamma_2 h}{g}-\left(f-\xi^{3/2}\right)^2 \propto c^2-\left(v-\frac{r}{t}\right)^2, \label{detM}
\end{equation}
where $c^2 = \gamma_2 p/\rho$ is the adiabatic sound speed in the post-shock fluid. {Here we used $\gamma_2$ instead of $\gamma$ to emphasize the fact that this expression remains valid even when one relaxes the constraint $\gamma_1 = \gamma_2 = 1+1/n$.} This demonstrates that there is a critical point within the flow, $r_{\rm c}$, which coincides with the sonic radius and satisfies

\begin{equation}
c^2(r_{\rm c})-\left(v(r_{\rm c})-\frac{r_{\rm c}}{t}\right)^2 = 0.
\end{equation}
Notice that, if the flow were time independent, this radius would reduce to the Bondi radius, where $v = c$ and the flow transitions from subsonic to supersonic. However, because of the outward motion of the shock, the sonic radius expands in time; nevertheless, the sonic radius does occur at a single value of the self-similar variable $\xi$, and we will denote this radius $\xi_{\rm c}$. 

\begin{figure}[htbp] 
   \centering
   \includegraphics[width=0.495\textwidth]{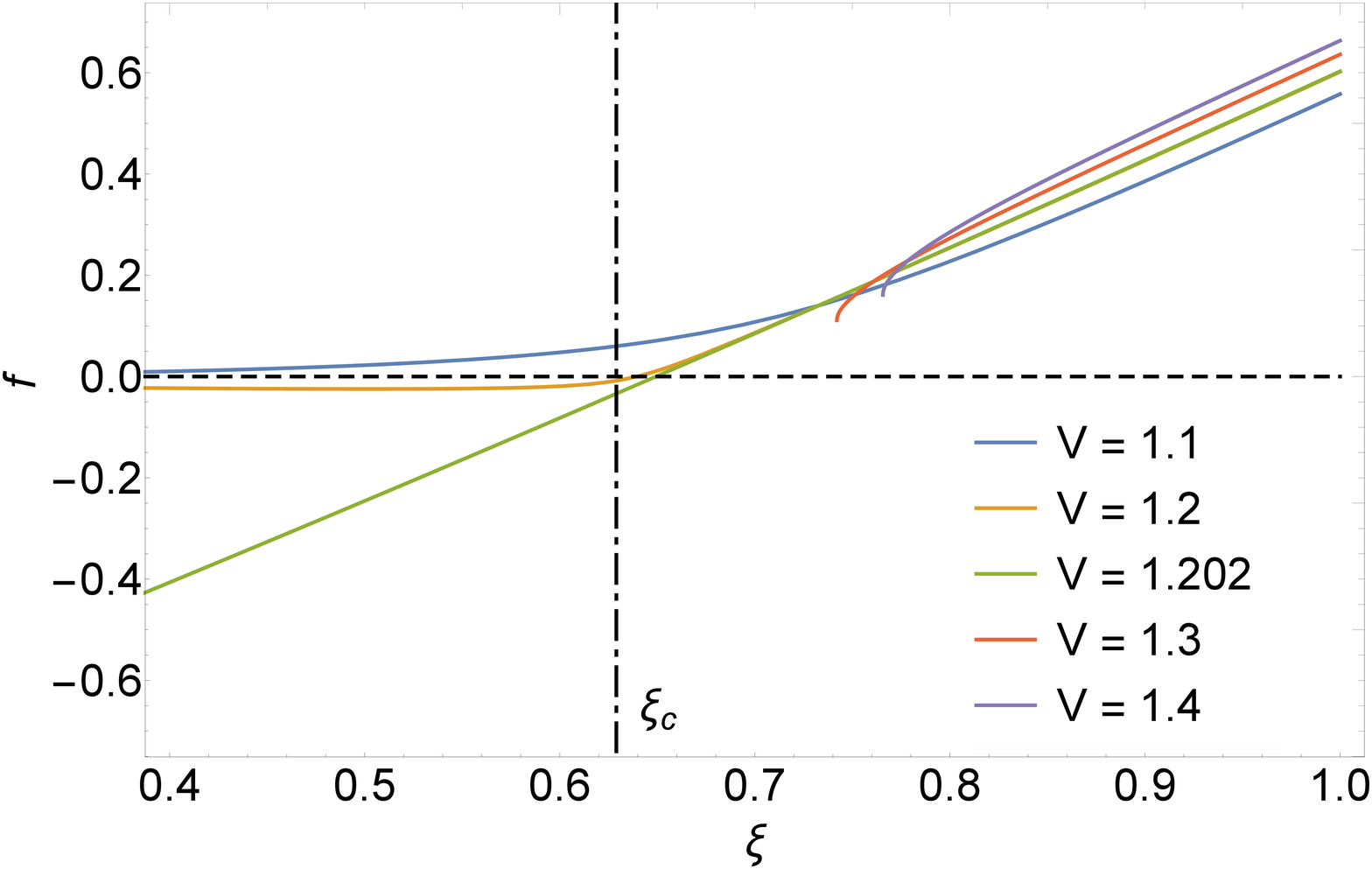} 
   \includegraphics[width=0.495\textwidth]{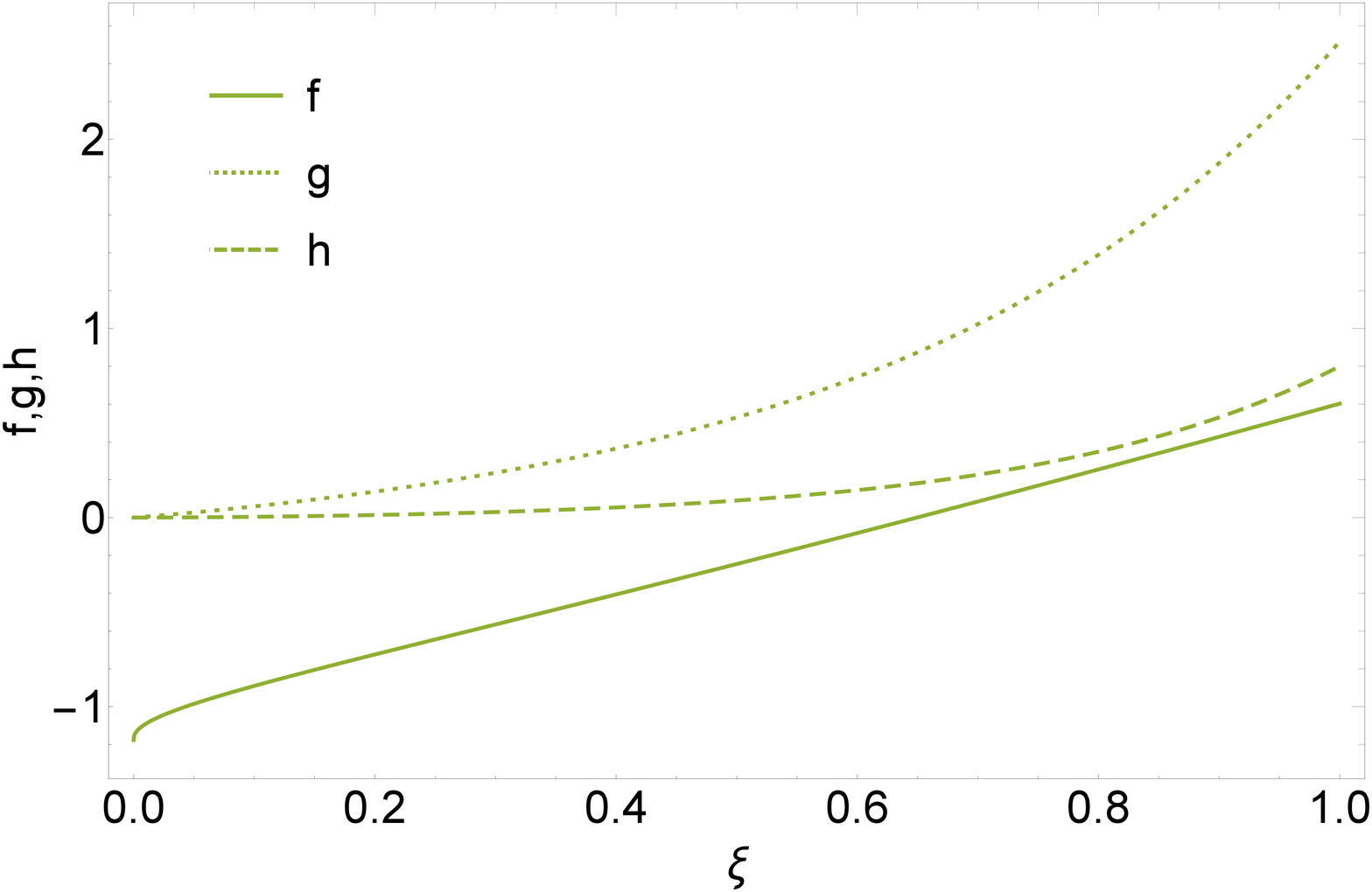} 
   \caption{Left: The function $f$, which is the velocity of the post-shock fluid normalized by the shock speed (which scales as the freefall speed), when $\gamma_1 = \gamma_2 = 1.4$ and $n = 2.5 = 1/(1.4-1)${}{; while the solutions are valid for general $\gamma_1$, $\gamma_2$, and $n$, one expects $\gamma_1 \simeq \gamma_2$ at these relatively small Mach numbers, and $n = 1/(\gamma_1 - 1)$ in the hydrogen envelopes of most massive stars, which was our motivation for choosing these values;} different curves correspond to different values of $V$, which sets the normalization of the shock speed; the vertical, dot-dashed line shows the value of the critcial point, $\xi_{\rm c} \simeq 0.629$, through which the solution smoothly passes when $V = V_{\rm c}$. Right: The self-similar functions for the velocity, density, and pressure, being $f$, $g$, and $h$, when $\gamma = 1.4$ and $V = V_{\rm c} = 1.202$. $V_{\rm c} = 1.202$ is the single value of the velocity, when $\gamma_1 = \gamma_2 = 1.4$ and $n=2.5$, that simultaneously satisfies the boundary conditions at the shock front and permits the smooth passage of the fluid quantities through the sonic point at $\xi_{\rm c} \simeq 0.629$}
   \label{fig:fofV}
\end{figure}

Taking the analogy with the Bondi problem further, we expect there to be only one solution (for fixed $\gamma_1$, $\gamma_2$, and $n$) that simultaneously satisfies the boundary conditions at the shock \emph{and} passes through the critical point. Because the smooth passage through the critical point places an additional constraint on the system, in general there will be only one value of $V$ (which we recall is the ratio of the shock velocity to the circular velocity) that smoothly connects the outward-moving shock to the origin through the sonic point. We find, indeed, that this is the case: The left-hand panel of Figure \ref{fig:fofV} shows the function $f$ -- the velocity of the fluid normalized by the shock velocity -- when $n=2.5$ and $\gamma = 1.4$ ($ = 1+1/n$). The different curves correspond to the values of $V$ shown in the legend. When $V < 1.202 \equiv V_{\rm c}$, the solutions remain subsonic, and the velocity approaches zero near the origin; when $V > V_{\rm c}$, the solutions becomes double valued for $\xi > \xi_c$, and, while mathematically possible, are physically untenable; and when $V = V_{\rm c}$, the solution passes through the critical point at $\xi_c = 0.629$ and the velocity approaches freefall near the origin (i.e., $f(0)\times V_c = -\sqrt{2}$). 

\begin{figure}[htbp] 
   \centering
   \includegraphics[width=0.495\textwidth]{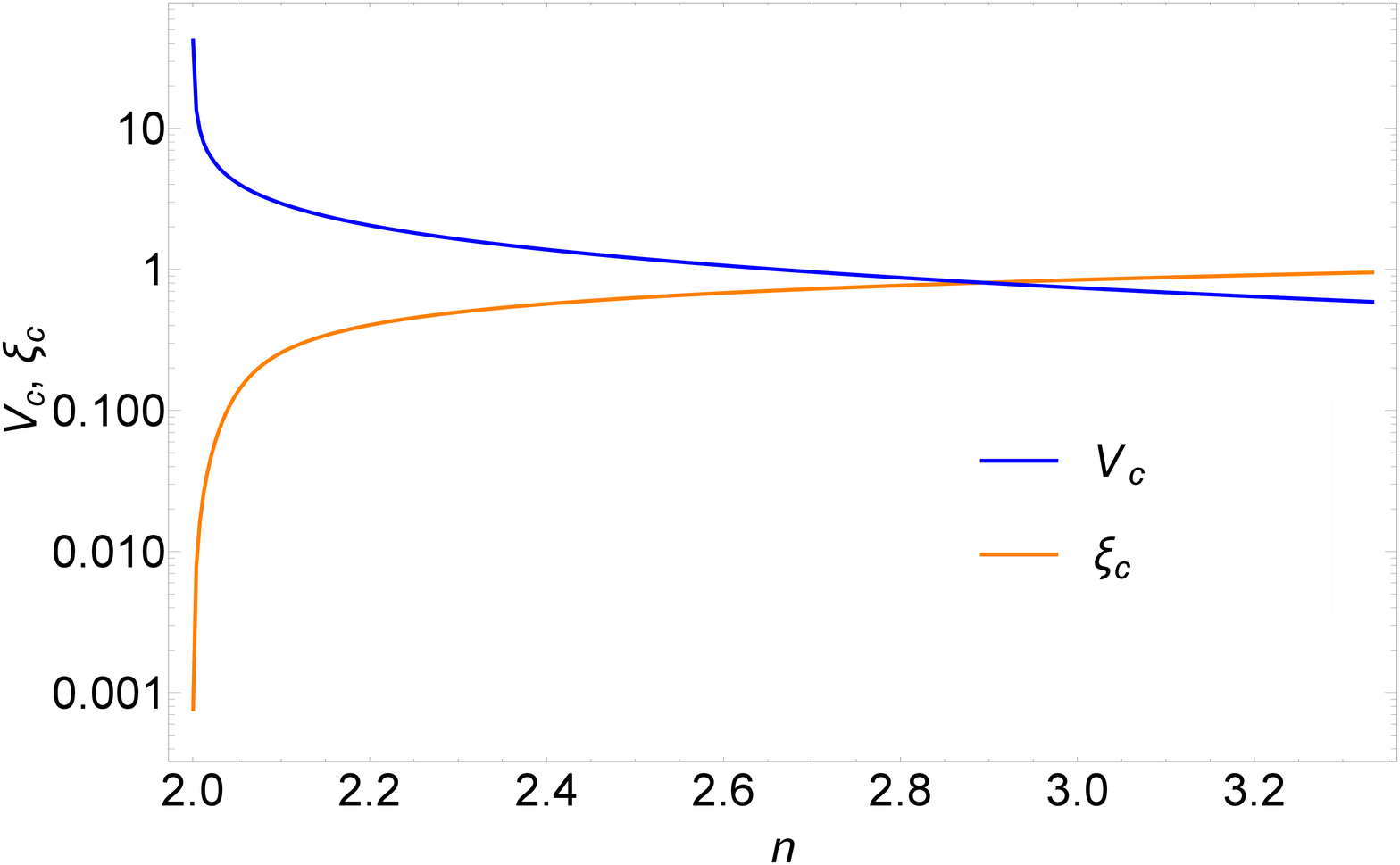} 
   \includegraphics[width=0.495\textwidth]{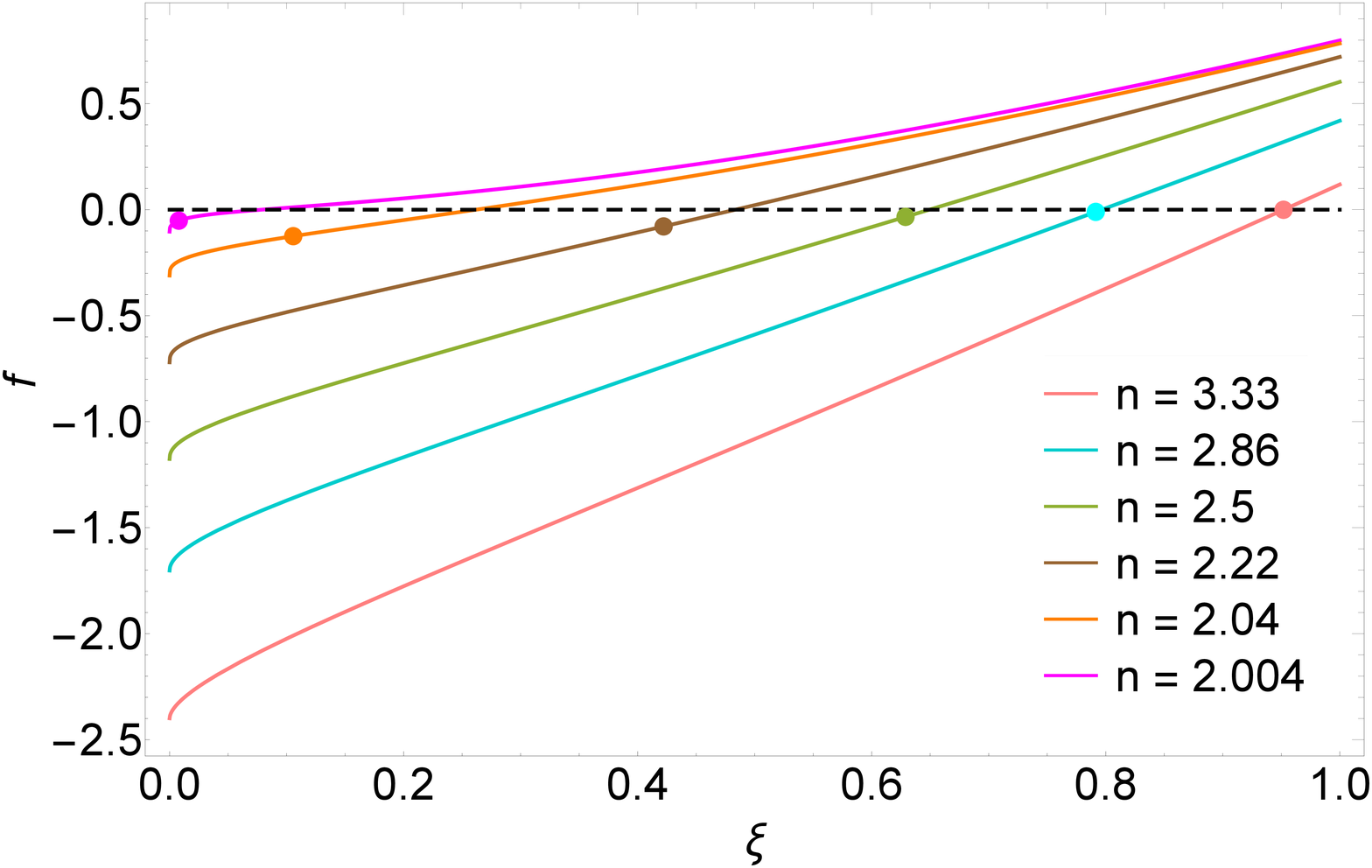} 
   \caption{ Left: The critical shock velocity, $V_{\rm c}$, for a transsonic solution and the sonic radius, $\xi_{\rm c}$, as functions of $n$ -- the power-law index of the ambient medium -- with $\gamma = 1+1/n$. Right: The normalized, post-shock velocity, given by the function $f$, that passes through the critical point for the values of $n$ shown in the legend, with $\gamma = 1+1/n$; the points on the curves illustrate the location of the sonic point, which is always within the region of the flow where the velocity becomes negative.}
   \label{fig:Vc}
\end{figure}

The solutions that avoid the critical point and remain subsonic to the origin are analogous to the ``settling'' solutions of the Bondi problem, and represent the situation in which the pressure gradient near the point mass is large enough to maintain approximate hydrostatic balance. While this configuration is, in principle, achievable, it is likely not physically realizable when the point mass is a black hole, as the pressure gradient needs to be infinitely large in order to resist the infall. A more likely outcome is that the pressure at some small radius falls below the necessary value to maintain hydrostatic balance, triggering accretion and causing the solutions to approach those that smoothly pass through the critical point. We therefore argue that the accreting solutions, which simultaneously satisfy the sonic point condition, are the physical ones, and it is these solutions that we primarily analyze for the remainder of this paper. 

The right-hand panel of Figure \ref{fig:fofV} illustrates the scaled velocity ($f$), density ($g$), and pressure ($h$) for the solution that passes through the critical point when $n = 2.5$ and $\gamma = 1.4$. While it may appear from the self-similar functions that the pressure and density approach zero near the origin, this is not necessarily the case, as $\rho \propto r^{-n}g(\xi) \propto \xi^{-n}g(\xi)$ and $p \propto r^{-n-1}h(\xi) \propto \xi^{-n-1}h(\xi)$. In fact, since the velocity approaches the freefall speed for small $r$, we expect the solutions to conform to $\rho \propto r^{-3/2}$ and $p \propto r^{-3\gamma/2}$ near the origin. These scalings are indeed upheld near the black hole, and they can be seen to satisfy Equations \eqref{sseq2} and \eqref{sseq3} in the limit that $f \rightarrow constant$ and $\xi \rightarrow 0${}{; we show this explicitly in Figure \ref{fig:density}, which gives the physical scalings of the density and pressure for a specific stellar progenitor}. 

The left-hand panel of Figure \ref{fig:Vc} gives the critical shock velocity, $V_{\rm c}$, and the sonic radius, $\xi_{\rm c}$, as functions of $n$ when $\gamma_1 = \gamma_2 = 1+1/n$. This figure demonstrates that as $n \rightarrow 2$, the shock velocity necessary to maintain smooth passage through the sonic radius and accretion onto the black hole increases to arbitrarily large values. This finding agrees with our arguments in Section \ref{sec:scalings}, being that as $n \rightarrow 2$, the shock needs an infinite amount of energy to overcome the logarithmically-diverging gravitational energy of the ambient medium. 
For the case of $n = 2$, it then follows that the only self-consistent and self-similar solution in the absence of additional energy injection is the one where the kinetic energy far outweighs the gravitational potential energy and the shock propagates in complete ignorance of the point mass at the origin, which is the Sedov-Taylor blastwave in a $\rho \propto r^{-2}$ ambient medium. Taking the limit as $V \rightarrow \infty$ and setting $n=2$ in Equations \eqref{sseq1} -- \eqref{sseq3} does, indeed, yield the Sedov-Taylor blastwave solution. 

The right-hand panel of Figure \ref{fig:Vc} shows the function $f$ that passes through the sonic radius for different values of $n$ when $\gamma = 1+1/n$, and the points denote the location of the sonic radius within the flow. As $n \rightarrow 2$, the necessary velocity to pass through the critical point (which approaches the origin in the $n \rightarrow 2$ limit) increases asymptotically, and the boundary conditions at the shock front reduce to their usual, large-$\mathscr{M}$ values that are independent of $V$. The function $f$ therefore converges to the Sedov-Taylor solution in this limit. We see that, for all solutions that pass through the critical radius, regions of the flow immediately behind the shock have positive velocity and move outward with the shock. However, at a finite value of the self-similar variable that is slightly larger than the sonic radius, the flow stagnates and, interior to that radius, falls back to the black hole. 

As $n$ increases, more of the post-shock fluid has a negative velocity, and the sonic point approaches the location of the shock itself. Moreover, there is a finite value of $n$ at which $f(1) = 0$; when $\gamma = 1+1/n$, this condition is achieved simultaneously with $V_{\rm c}^2 = 1/n = c^2$, where $c^2$ is the sound speed of the ambient medium normalized by the circular speed. Therefore, in this limit, the shock reduces to a small perturbation that travels at the sound speed and simply informs the fluid of the accretion at the origin, and the velocity is negative everywhere (i.e., the shock does not induce any outward motion). This case is most similar to Bondi accretion, as there is no outward motion anywhere within the fluid. However, the solution is still time dependent, as infall is initiated on the sound crossing time from the origin, and the sonic point no longer lies within the flow but coincides with the location of the shock front. 

{We find that $n =3.5$ is the value of $n$, with $\gamma_1 = \gamma_2 = 1+1/n = 9/7$, above which there are no solutions with a shock of finite strength (i.e., such that the post-shock velocity is greater than zero). This behavior likely stems from the fact that the energy density of the ambient medium falls off as a steeper function of radius as $n$ increases. There is then a point where, in order to satisfy the both the critical point condition and the jump conditions, the shock would need to have a subsonic velocity in order to meet the energy constraint. This then implies that, when the density falls off faster than $\rho \propto r^{-3.5}$, any shock of finite magnitude will be forced to accelerate and fall outside this regime of self-similarity.}

\subsection{Energy}
Multiplying Equation \eqref{rmom} by $\rho r^2 v$ and performing some simple manipulations yields the energy equation:

\begin{equation}
\frac{\partial}{\partial t}\left[\mathscr{E}\rho r^2\right]+\frac{\partial}{\partial r}\left[\mathscr{B}\rho r^2 v\right] = 0, \label{energy}
\end{equation}
where

\begin{equation}
\mathscr{E} = \frac{1}{2}v^2+n\frac{p}{\rho}-\frac{GM}{r} = \frac{GM}{r}\left(\frac{1}{2}V^2f^2+nV^2\frac{h}{g}-1\right) \label{eden}
\end{equation}
and

\begin{equation}
\mathscr{B} = \mathscr{E}+\frac{p}{\rho} = \frac{1}{2}v^2+(n+1)\frac{p}{\rho}-\frac{GM}{r} = \frac{GM}{r}\left(\frac{1}{2}V^2f^2+(n+1)V^2\frac{h}{g}-1\right) \label{bden}
\end{equation}
are the specific energy and the Bernoulli parameter of the fluid. Integrating the energy equation from 0 to $r_{\rm sh}(t)+\epsilon$ with $\epsilon$ a positive number, performing some rearrangements, and taking the limit as $\epsilon \rightarrow 0$ gives\footnote{One could also let $\epsilon$ be a negative number and use the relevant quantities for post-shock fluid; the result is identical to the one here owing to the conservation of energy and the jump conditions.}

\begin{equation}
\frac{\partial}{\partial t}\int_0^{r_{\rm sh}(t)}4\pi\mathscr{E}\rho r^2dr = \frac{\partial E_{\rm sh}}{\partial t} = 4\pi\rho_{\rm 1}v_{\rm sh}r_{\rm sh}^2\mathscr{E}_1+\lim_{r\rightarrow 0}\left(4\pi \rho vr^2\mathscr{B}\right), \label{totenergy}
\end{equation}
where $E_{\rm sh}$ is the total energy contained in the shock, $\rho_{\rm 1}$ and $\mathscr{E}_{\rm 1}$ are the mass density and the energy density of the ambient medium, and we introduced the factor of $4\pi$ to account for the integral of the energy over the solid angle. As we argued in Section \ref{sec:scalings}, we see that the energy contained in the shocked fluid is modified by the binding energy of the ambient medium that is swept up by the shock, and by the flux of energy that is drained onto the black hole. 

Changing variables from $r$ to $\xi$, Equation \eqref{totenergy} becomes

\begin{equation}
\frac{\partial E_{\rm sh}}{\partial t} = 4\pi\rho_0 r_0^2 v_{\rm sh}\frac{GM}{r_0}\left(\frac{r_{\rm sh}}{r_0}\right)^{1-n}\left\{-\frac{1}{n+1}+\lim_{\xi\rightarrow 0}\left[\xi^{\frac{1}{2}-n}\left(\frac{1}{2}V^2f^2+V^2(n+1)\frac{h}{g}-1\right)f g\right]\right\} \label{energytot}
\end{equation}
The first term in braces on the right-hand side is the binding energy of the ambient medium, and reduces the total energy contained in the flow. The second term represents the accretion of energy onto the black hole; interestingly, because the fluid is bound to the black hole prior to being accreted in all of our solutions (see Figure \ref{fig:ebofr}), the accretion term actually represents a \emph{source} of energy for the shocked fluid -- accreting material removes negative energy from the flow. \footnote{{}{While it is true that accretion onto the black hole adds energy to the flow, we note that the existence of the sonic point implies that any changes to that accretion rate are causally disconnected from the fluid immediately behind the shock. Therefore, it is more accurate to state that the accretion of energy \emph{through the sonic radius} (which is capable of influencing both the material in the vicinity of the black hole and the shock) is responsible for directly modifying the dynamics of the post-shock fluid and the propagation of the shock itself.}}

\begin{figure}[htbp] 
   \centering
   \includegraphics[width=0.995\textwidth]{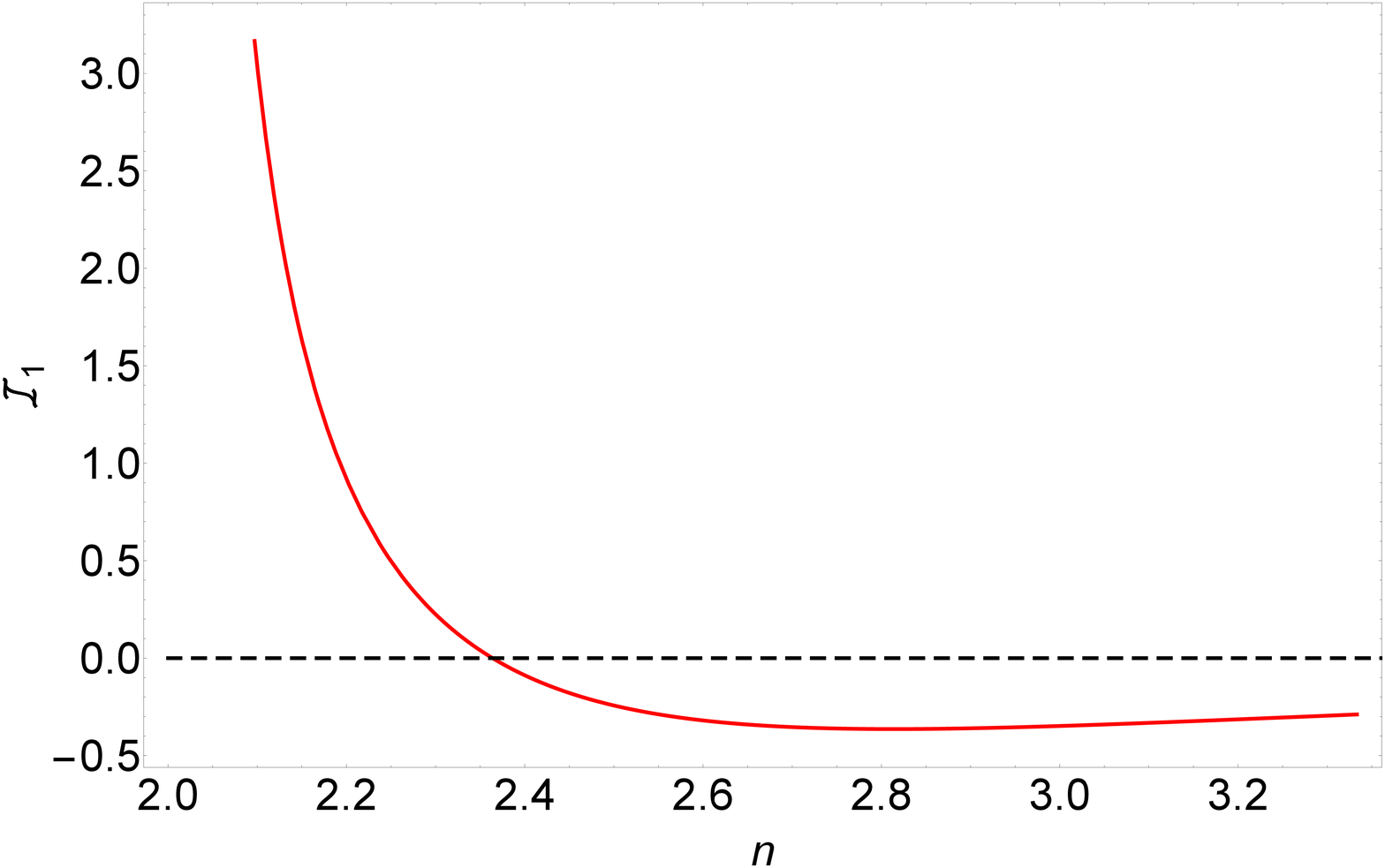} 
   \caption{The integral $I_1(n)$ appearing in Equation \eqref{Etot}, which gives the total energy contained in the shocked fluid, as a function of the power-law index of the density profile of the ambient medium, $n$. While this figure shows that this integral is typically a number of order unity, there is a special value of $n$, equal to $\simeq 2.36$, at which the energy contained in the blastwave is identically zero.}
   \label{fig:Iofg}
\end{figure}

\begin{figure}[htbp] 
   \centering
   \includegraphics[width=0.495\textwidth]{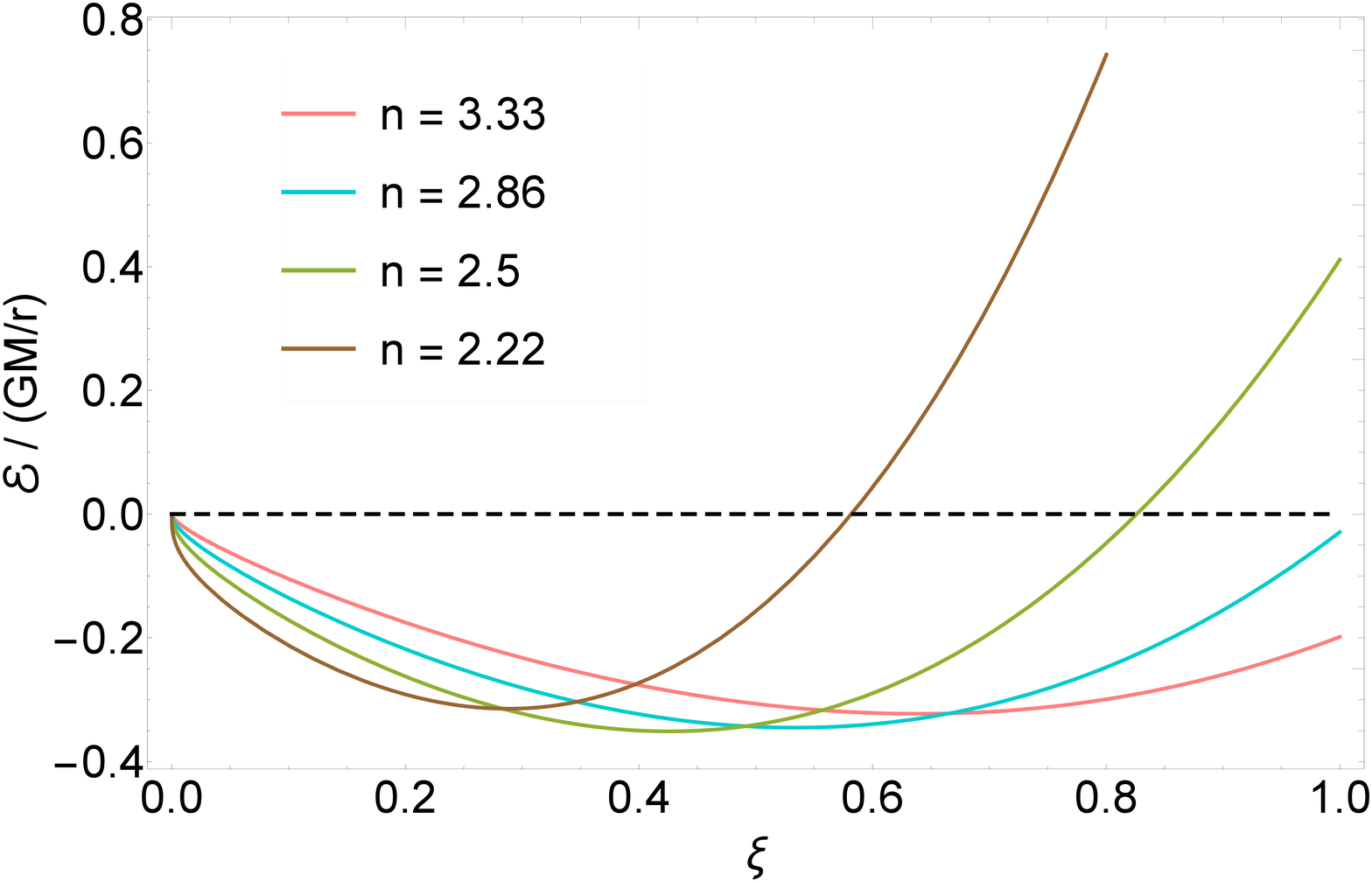} 
   \includegraphics[width=0.495\textwidth]{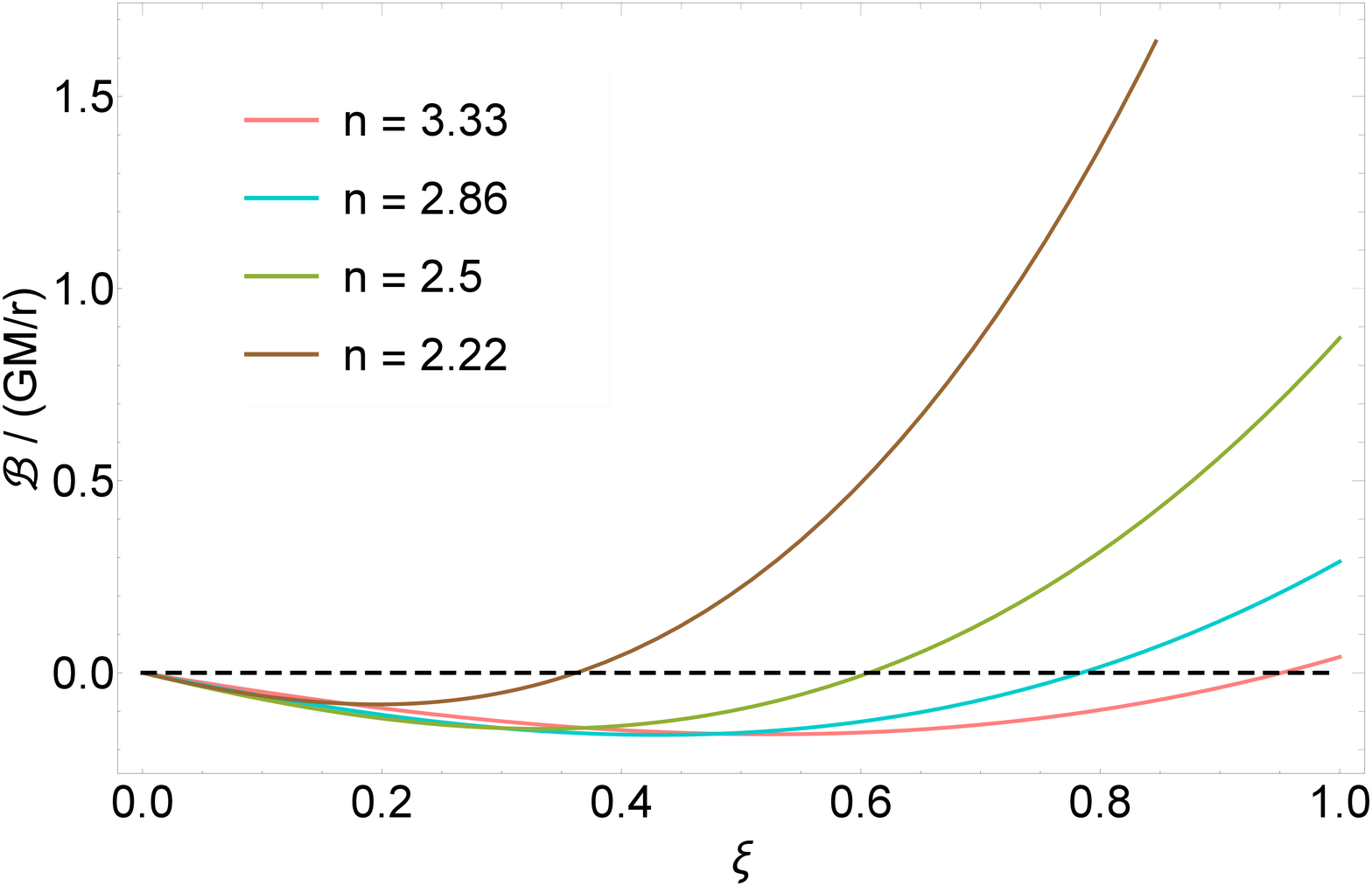}
   \caption{Left: The self-similar energy density of the fluid as a function of $\xi$, given by Equation \eqref{eden}, for the values of $n$ in the legend, where $n$ is the power-law index of the density profile of the ambient medium. Right: The self-similar Bernoulli parameter of the fluid, given by Equation \eqref{bden}, for the same values of $n$ as the left-hand panel.}
   \label{fig:ebofr}
\end{figure}

We can also change variables from $r$ to $\xi$ in the left-hand side of Equation \eqref{energytot} {}{(i.e., under the integral that defines the energy of the shock in Equation \ref{totenergy})}, which gives

\begin{equation}
\frac{\partial E_{\rm sh}}{\partial t} = -(n-2)4\pi\rho_0r_0^2v_{\rm sh}\frac{GM}{r_0}\left(\frac{r_{\rm sh}(t)}{r_0}\right)^{1-n}\int_0^{1}\xi^{1-n}\left(\frac{1}{2}V^2f^2+nV^2\frac{h}{g}-1\right)g\,d\xi \label{Etot2}
\end{equation}
Comparing this expression with Equation \eqref{energytot} then yields an identity relating an integral of the self-similar functions to the boundary conditions; in either case, we see that the energy of the shocked fluid evolves according to

\begin{equation}
E_{\rm sh} = 4\pi \rho_0 r_0^3\frac{GM}{r_0}\left(\frac{r_{\rm sh}}{r_0}\right)^{2-n}\int_0^{1}\xi^{1-n}\left(\frac{1}{2}V^2f^2+nV^2\frac{h}{g}-1\right)g\,d\xi \equiv 4\pi \rho_0 r_0^3\frac{GM}{r_0}\left(\frac{r_{\rm sh}}{r_0}\right)^{2-n}{I}_1(n) \propto t^{\frac{4}{3}-\frac{2n}{3}} \label{Etot}.
\end{equation}
The temporal evolution of the energy, given by the last line of this expression, agrees with the heuristic arguments of Section \ref{sec:scalings}. When $n = 2$, the energy is independent of time, which is consistent with the fact that this limit corresponds to the Sedov-Taylor blastwave (which is energy-conserving by assumption). When $n>2$, the energy is a decreasing function of time, demonstrating that the energy of the shocked fluid is not conserved; this finding is in agreement with our expectations from Section \ref{sec:scalings}, given that the shock is sweeping up the binding energy of the ambient medium and the black hole is accreting bound material. Furthermore, the \emph{sign} of the integral in Equation \eqref{Etot} depends on $n$: Figure \ref{fig:Iofg} shows the integral in Equation \eqref{Etot} as a function of $n$. For $n \lesssim 2.36$, the integral is negative, which implies that the flux of bound material onto the black hole is removing negative energy at a faster rate than that at which the shock is sweeping up bound material (i.e., the \emph{derivative} is positive, so the overall energy is negative and the rate of change of the energy is increasing with time). For $n \gtrsim 2.36$, the integral is positive and the addition of bound material outweighs the subtraction from accretion, so the rate of change of the energy is negative as more bound material is added to the flow. And, finally, there is a power-law index $n \simeq 2.36$ at which the contributions from the sweeping up of bound material and the accretion of material \emph{exactly balance}, and the total energy of the shocked fluid is conserved at zero. 

The left-hand panel of Figure \ref{fig:ebofr} illustrates the self-similar specific energy (i.e., Equation \ref{eden} normalized by $GM/r$), and the right-hand panel gives the self-similar Bernoulli parameter (Equation \ref{bden} normalized by $GM/r$), both as functions of $\xi$ and for the $n$ shown in the legend. We see that for $n \gtrsim 2.86$, the specific energy of the entire fluid behind the shock is negative, but as $n$ decreases there is a region of positive energy that extends to modest values of $\xi$; this is consistent with the fact that there is a value of $n$ at which the total, integrated energy vanishes identically. On the other hand, for $n < 3.5$ there is always at least some region in the vicinity of the shock that has a positive Bernoulli parameter, and when $n = 3.5$ (when the post-shock velocity is zero and the system is still in hydrostatic equilibrium at the shock front) the Bernoulli parameter is exactly zero at the location of the shock and is negative for all $\xi < 1$. 

\subsubsection{Lagrangian evolution}
\label{sec:lagrangian}
The positivity of the Bernoulli parameter and, for smaller $n$, the internal energy near the shock naively suggests that some fraction of the gas is unbound from the system and will escape to infinity. To determine whether or not this is the case, consider the Lagrangian evolution of a fluid element, governed by the equation

\begin{equation}
\frac{dr}{dt} = v = V\sqrt{\frac{GM}{r}}f\left(\frac{r}{r_{\rm sh}(t)}\right).
\end{equation}
Let the shock satisfy $r_{\rm sh}(T) = R$, where $T = 2R^{3/2}/(3V\sqrt{GM})$, and define $u \equiv r/R$ and $\tau \equiv t/T$; the previous equation then becomes

\begin{equation}
\frac{du}{d\tau} = \frac{2}{3}u^{-1/2}f\left(\frac{u}{\tau^{2/3}}\right), \label{lag}
\end{equation}
and the initial condition is $u(1) = u_0$, with $u_0 \le 1$. The limit $u(\tau\rightarrow \infty)$ then gives the asymptotic evolution of the fluid parcels.

Since the shock position scales as $r_{\rm sh} \propto \tau^{2/3}$, gas parcels behind the shock can at most follow orbits that satisfy $u \propto \tau^{2/3}$, as otherwise the fluid element would overtake the shock. On the other hand, if the position scales as a weaker power of time than $\tau^{2/3}$, then eventually the ratio $u/\tau^{2/3}$ will fall below the value of $\xi$ at which the velocity equals zero, and the fluid element will fall back to the origin. Therefore, \emph{if any portion of the post-shock fluid is unbound}, then it must asymptotically follow $u = \xi_0\tau^{2/3}$ with $\xi_0$ a constant less than one. Inserting this expression into Equation \eqref{lag} gives

\begin{equation}
\xi_0^{3/2} = f(\xi_0), \label{Ccrit}
\end{equation}
which is an algebraic condition that yields the value(s) of $\xi_0$ (related to the initial position of the fluid element) for which fluid elements are unbound. However, it is apparent from Figure \ref{fig:Vc} that this condition is not satisfied for any of the self-similar solutions. We thus conclude that there cannot be any physical solutions for which Equation \eqref{Ccrit} is satisfied, and hence the entire, post-shock gas must be bound to the black hole. \footnote{An exception is when $n = 2$ and the Sedov-Taylor blastwave is the solution, and $f(\xi_0) = \xi_0^{3/2}$ \emph{at} the origin. All fluid elements asymptotically follow marginally-bound orbits in this case.} 

\begin{figure}[htbp] 
   \centering
   \includegraphics[width=0.495\textwidth]{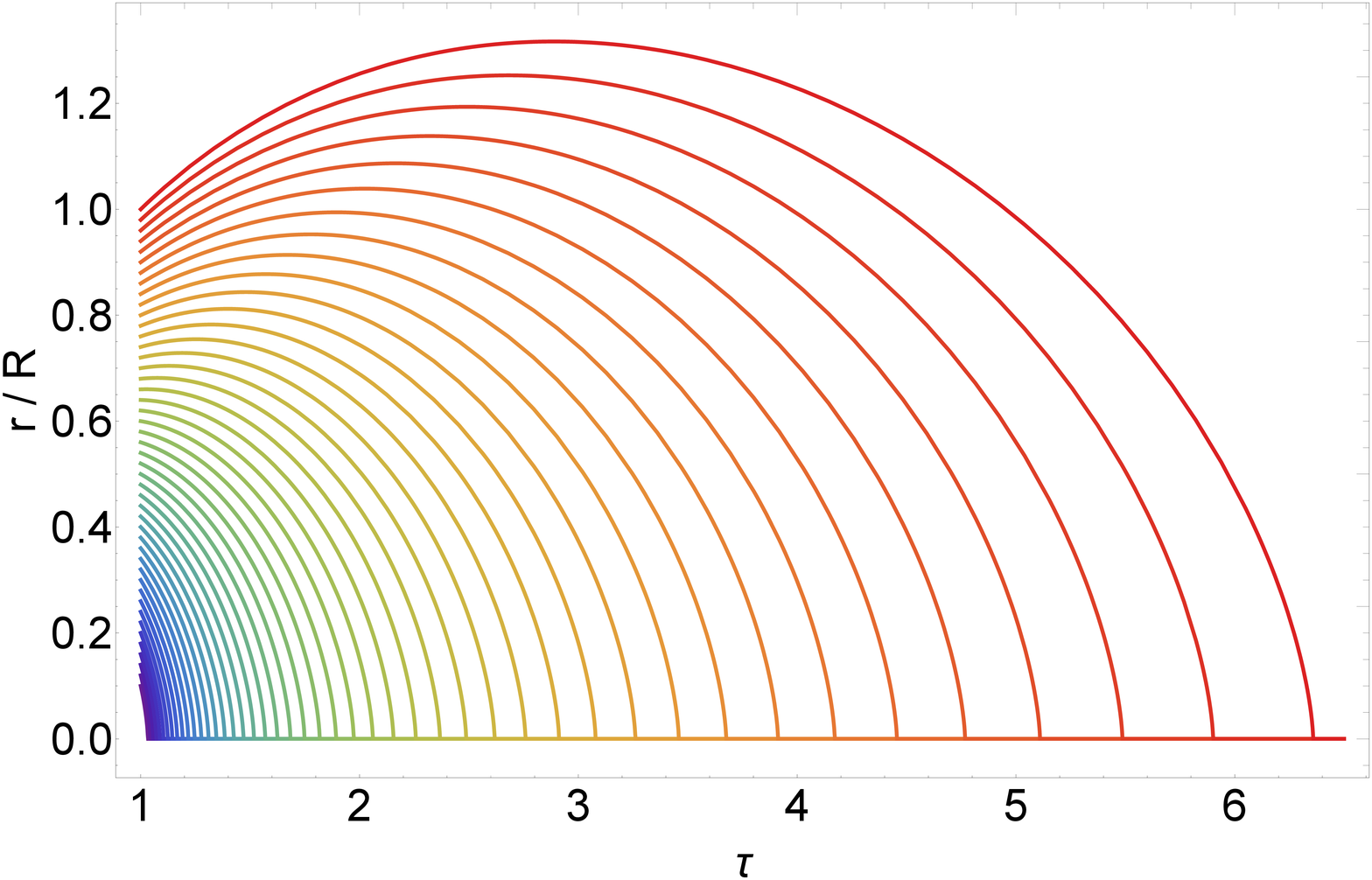} 
   \includegraphics[width=0.495\textwidth]{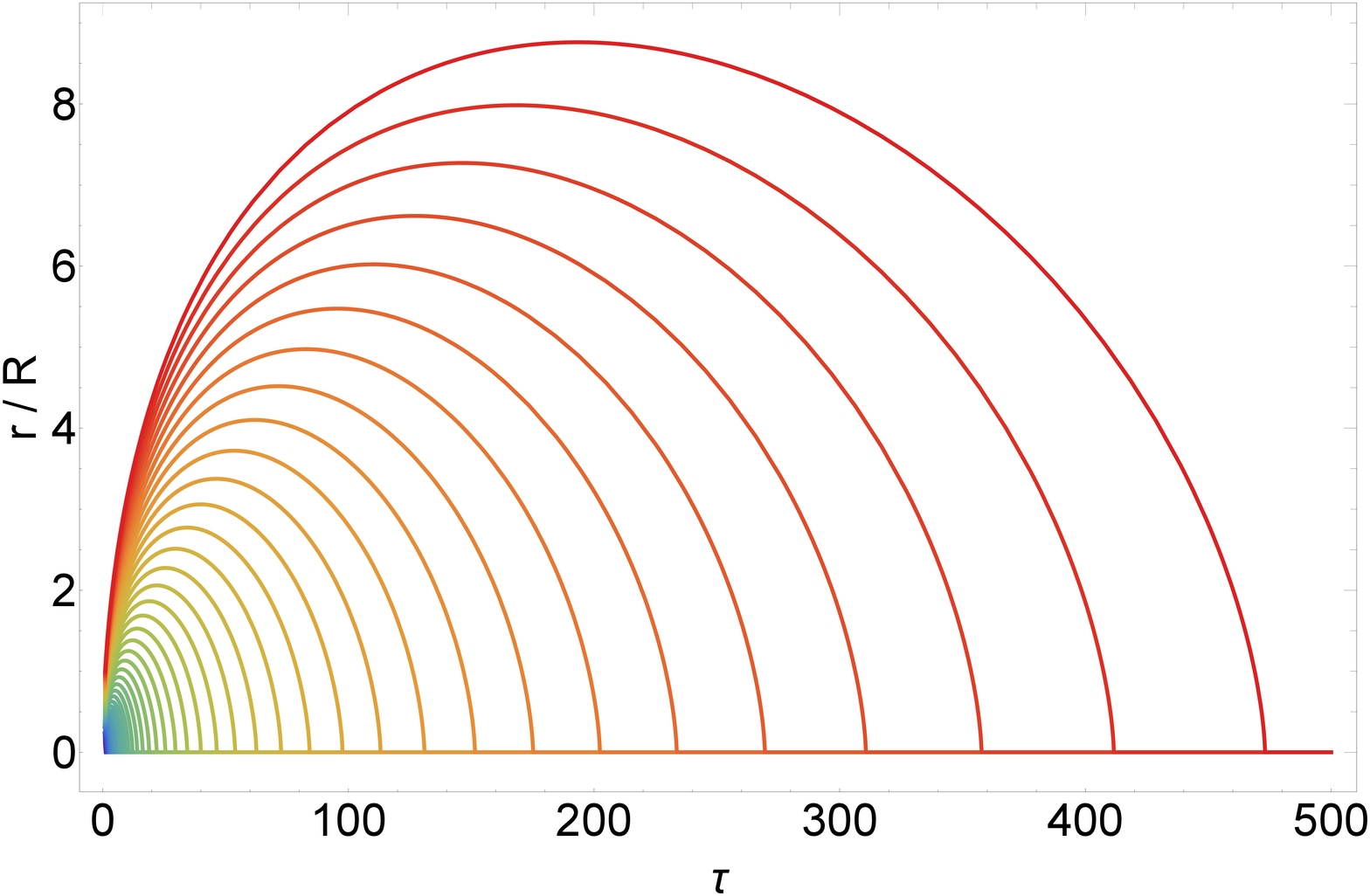} 
   \caption{The Lagrangian positions of the fluid elements as functions of $\tau = t/T$, where $T = 2R^{3/2}/(3V\sqrt{GM})$ and $R$ is the location of the shock at $\tau = 1$, when $n = 2.5$ (left) and $n = 2.04$ (right), $n$ being the power-law index of the density profile of the ambient medium; the fluid elements are color-coded by their initial distance behind the shock for clarity. This figure illustrates that, even when fluid elements have a large energy and Bernoulli parameter immediately behind the shock (see Figure \ref{fig:ebofr}), every gas parcel is bound and eventually returns to the origin in a finite time.}
   \label{fig:riplot}
\end{figure}

In support of this argument, Figure \ref{fig:riplot} shows the Lagrangian evolution of the fluid parcels behind the shock when $n = 2.5$ (left panel) and $n=2.04$ (right panel), obtained by solving Equation \eqref{lag} numerically for a number of different initial positions behind the shock (i.e., the value of $r/R$ at $\tau = 1$). The colors simply serve to distinguish between different fluid elements, which are at different post-shock positions at $\tau = 1$. We see that, in agreement with the solution for the self-similar velocity profile, gas parcels closer to the shock front have a larger initial velocity and more of the solution has positive velocity initially for $n = 2.04$. Nonetheless, every fluid element eventually reaches a relative maximum and falls back to the origin in a finite amount of time. When $n = 2.5$, the maximum distance is only slightly greater than the initial shock radius, but for $n = 2.04$, gas parcels extend out to nearly ten times the initial shock position and the time taken to fall to the origin is two orders of magnitude larger than the freefall time from $R$. Moreover, the internal energy and Bernoulli parameter of the gas for $n = 2.04$ are much greater than zero (and marginally greater than zero when $n = 2.5$; see Figure \ref{fig:ebofr}), yet the fluid parcels remain bound to the hole; this shows that simply having a positive energy -- even a very large one -- does not necessarily guarantee that the gas escapes to infinity.

\subsection{Mass and Accretion Rate}
Multiplying the continuity equation \eqref{cont} by $4\pi r^2$, integrating from $0$ to $r_{\rm sh}(t)+\epsilon$ and performing the same sequence of operations as we did for the energy equation gives

\begin{equation}
\frac{\partial}{\partial t}\int_0^{r_{\rm sh}(t)}4\pi\rho r^2 dr \equiv \frac{\partial M_{\rm sh}}{\partial t}= 4\pi\rho_0r_0^2\left(\frac{r_{\rm sh}(t)}{r_0}\right)^{2-n}v_{\rm sh}\left\{1+\lim_{\xi\rightarrow 0}\left(fg\xi^{\frac{3}{2}-n}\right)\right\}, \label{mdot}
\end{equation}
where $M_{\rm sh}$ is the mass contained in the shocked fluid. As was true for the energy, there are two boundary terms that contribute to the rate of change of the mass of the shocked fluid, the first term in braces being the mass swept up from the ambient medium, while the second is due to the accretion of material onto the black hole. Again, writing the left-hand side in terms of self-similar variables yields an integral identity for the self-similar functions, and we find that the mass of the shocked fluid evolves as

\begin{equation}
M_{\rm sh} = 4\pi \rho_0 r_0^3\left(\frac{r_{\rm sh}(t)}{r_0}\right)^{3-n}\int_0^{1}\xi^{2-n}gd\xi \propto t^{2-\frac{2n}{3}}.
\end{equation}
Unlike the energy of the shock, the sign of the integrand, and hence the integral itself, is always positive here, as is the exponent for all $n <3$. This finding demonstrates that even though the black hole is consuming material from the flow, the added mass from the ambient medium outweighs the accretion, resulting in a net increase in the mass of the shocked fluid as time advances. 

Equation \eqref{mdot} shows that the accretion rate onto the black hole is given by

\begin{equation}
\dot{M}_{\bullet} = -4\pi \rho_0 r_0^2 V\sqrt{\frac{GM}{r_{\rm sh}}}\left(\frac{r_{\rm sh}}{r_0}\right)^{2-n}\lim_{\xi \rightarrow 0}\left(fg\xi^{\frac{3}{2}-n}\right) \propto t^{1-\frac{2n}{3}}. \label{mdotbh}
\end{equation}
This power law is negative for all $n < 3/2$, and varies from $\dot{M}_{\bullet} \propto t^{-1}$ for $n = 3$ to $\dot{M}_{\bullet} = t^{-1/3}$ for $n = 2$. These power-law fallback rates are significantly shallower than the prediction $\dot{M}_{\bullet} \propto t^{-5/3}$, which occurs when the shock has enough energy to unbind material from the system, leaving only very weakly bound material to return to the black hole at late times \citep{chevalier89}. This temporal scaling is also the same that one would expect if the gas were in pure freefall, which results from the scaling of the shock velocity and is in agreement with the heuristic arguments put forth in Section \ref{sec:scalings}.

\subsection{Wind-like solutions}
\label{sec:winds}
The gravitational energy of the ambient medium diverges logarithmically when $\rho \propto r^{-2}$, and diverges more strongly when the density falls off less rapidly than $\rho \propto r^{-2}$. In these cases, self-similar solutions with finite energy and accretion at the center cannot be maintained, as -- from Figure \ref{fig:ebofr} -- the black hole only removes negative energy from the system. As time advances, however, the shock needs more and more \emph{positive} energy in order to maintain the same, $r_{\rm sh} \propto t^{2/3}$ scaling, and the best situation that can be actualized is that the black hole accretes all of the negative energy from the post-shock fluid.  The shock would therefore need an infinite velocity, and therefore add an infinite amount of energy to the post-shock gas, in order to simultaneously propagate to infinity and yield accretion at the center when the density falls off shallower than $\rho \propto r^{-2}$. 

\begin{figure}[htbp] 
   \centering
   \includegraphics[width=0.495\textwidth]{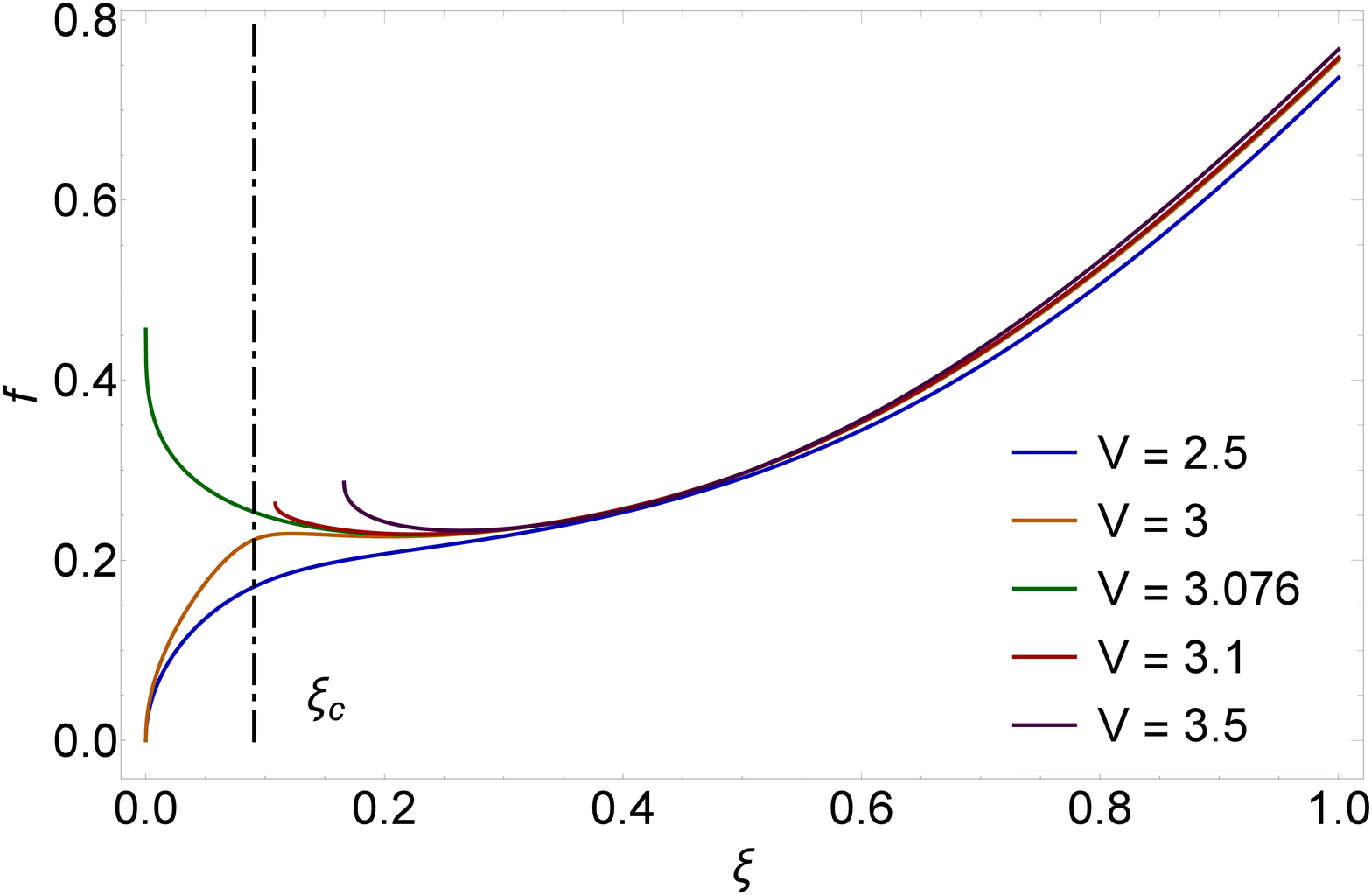} 
    \includegraphics[width=0.495\textwidth]{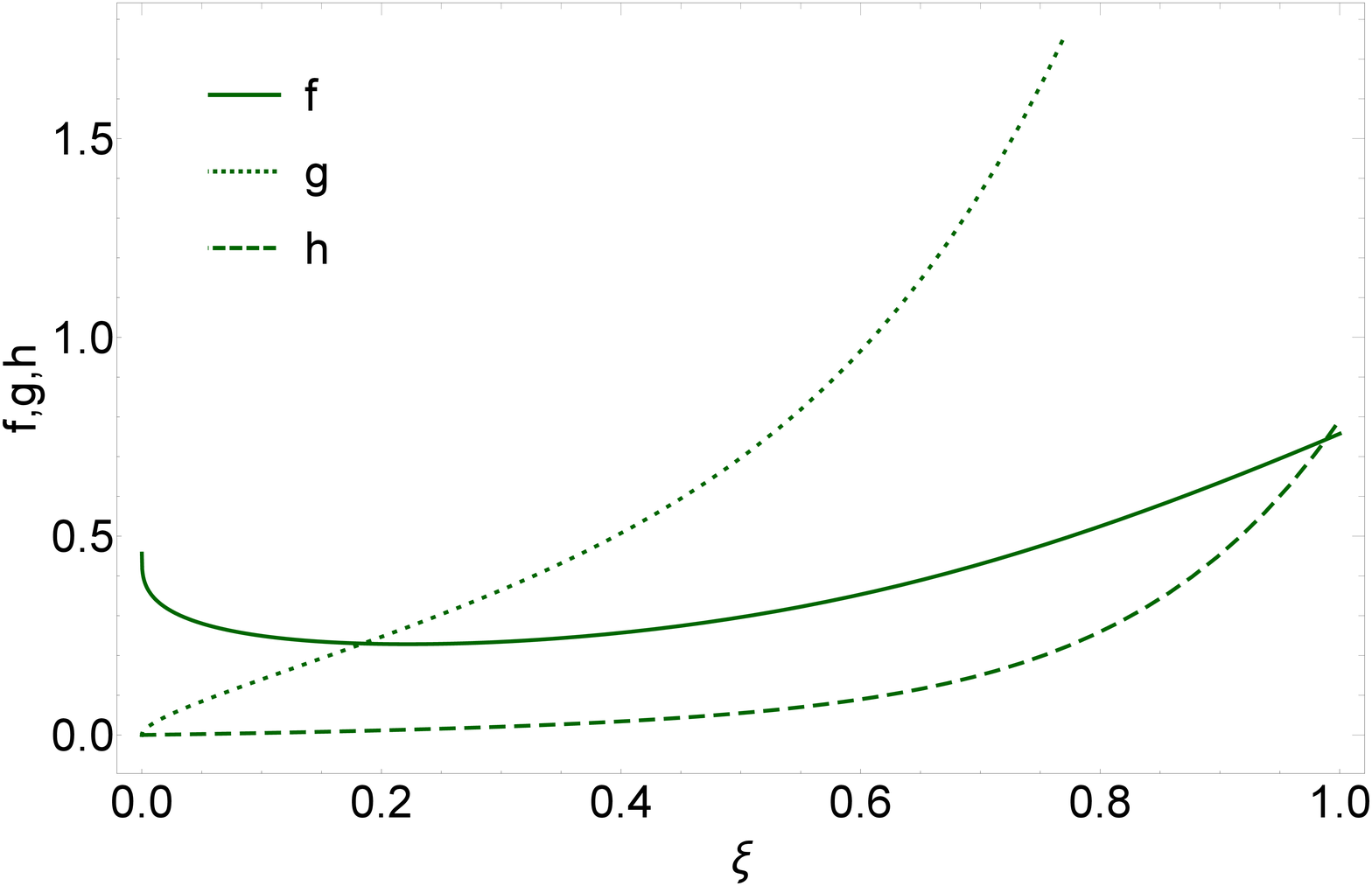} 
   \caption{Left: The self-similar function for the velocity, $f$, when the density falls off as $\rho \propto r^{-2}$, $\gamma = 1.5$, and the dimensionless velocity of the shock takes on the values shown in the legend; in agreement with the accreting solutions, there is one value of the shock velocity, being $V_{\rm c} \simeq 3.076$, that permits the solution to transition from super to sub-sonic through the sonic radius $\xi_{\rm c} \simeq 0.0907$ and has a velocity that scales as the escape velocity near the origin. Solutions with shock velocities below this critical value approach a constant velocity near the origin, as the function $f$ scales as $f \propto \xi^{1/2}$ for $\xi \ll 1$, and solutions with $V$ greater than this value are double valued. Right: The solutions for the dimensionless velocity ($f$), density ($g$), and pressure ($h$) that pass through the critical point for $n = 2$ and have a positive energy flux near the origin. These solutions constitute the analogue of the Parker wind when the outer boundary condition is a shock that moves at $r_{\rm sh} \propto t^{2/3}$.}
   \label{fig:parker}
\end{figure}

There is, however, a separate ``class'' of solution when $\rho$ falls off shallower than $\rho \propto r^{-2}$ that have \emph{outflow} near the origin instead of accretion. As for the accreting solutions, we find that for $\gamma = 1+1/n$ there is only one value of the shock velocity that permits passage through a sonic point within the flow. When $n = 2$, we find that the critical velocity is $V_{\rm c} \simeq 3.016$, and the left-hand panel of Figure \ref{fig:parker} shows how the solutions for $f$ change as we vary the shock velocity: when $V < V_{\rm c}$, the self-similar function $f$ scales as $f \propto \xi^{1/2}$ near the origin, and the velocity approaches a constant; when $V > V_{\rm c}$, the velocity becomes double-valued outside of a certain point within the flow; and when $V = V_{\rm c}$, the solution passes through a critical point at $\xi_{\rm c} \simeq 0.0907$, and the self-similar function $f$ equals a finite value near the origin. The right-hand panel of Figure \ref{fig:parker} illustrates the self-similar functions for the velocity ($f$), density ($g$), and pressure ($h$) for the solution that passes through the critical point; as was true for the solutions with accretion, the density and pressure approach $\rho \propto r^{-3/2}$ and $p \propto r^{-5/2}$ near the origin in this case.

We interpret these solutions as being analogous to the Parker wind when the critical point condition is satisfied, and those with $V < V_{\rm c}$ correspond to ``breezes'' that remain subsonic. Importantly, this class of solutions only exists because there is a source of energy at the origin, and when $n = 2$ the rate of energy injection exactly counterbalances the rate at which the shock is sweeping up the binding energy of the ambient medium. When $n > 2$, the energy injection rate outweighs the consumption of binding energy from the ambient medium, and the overall energy contained within the shocked fluid grows in time. 

Unlike the accretion solutions, which displayed qualitatively similar behavior when the adiabatic indices were not equal to the polytropic index of the ambient medium, this is not the case for the wind solutions, and when $\gamma_1 = \gamma_2 \neq 1+1/n$ we find that there is generally only a finite radius into which the solutions extend from the shock before the density diverges. The reason for this is that, if we rearrange Equation \eqref{sseq3}, then we find

\begin{equation}
\xi\left(f-\xi^{3/2}\right)\frac{d}{d\xi}\ln\left(\frac{h}{g^{\gamma_2}}\right) = \frac{\gamma_3-\gamma_2}{\gamma_3-1}f,
\end{equation}
where $\gamma_3 = 1+1/n$. Writing the entropy equation in this way shows that, if $f = \xi^{3/2}$ anywhere within the fluid, then the entropy of the gas must diverge at that point \emph{unless $\gamma_2 = \gamma_3$ or $f(\xi) = \xi^{3/2} = 0$}. Written in terms of physical coordinates, $f-\xi^{3/2} \propto v-r/t$, which shows that this location within the solution corresponds to a contact discontinuity between the inner flow launched from the point mass and the shocked fluid. 

While these solutions could be realized in nature, it is difficult to see how they could be sustained, as the energy injection from the point source must be exactly right to balance the inward flux of gravitational energy from the ambient medium. However, the physical processes dictating the energy generation (e.g., a disk wind or pulsar spin-down energy) are local, and it is difficult to see how the point source could self-regulate to the point where this precise matching could be achieved. At any rate, the problems considered in this paper are those for which matter is accreted, not generated, from the origin, and hence we do not consider these Parker wind analogues further.

\subsection{Comparison to previous works}
Before moving on to the application of the previous models to failed supernovae, here we briefly compare and contrast our results to the two other, relatively recent investigations of similar phenomena of which we are aware. The first is \citet{chevalier89}, who proposed a self-similar model for the propagation of a shock in the gravitational field of a neutron star, and restricted the analysis to the case in which the density profile of the ambient medium is described by the power law $\rho \propto r^{-2}$ (his Section 2). The motivation for considering this special profile of the ambient density is that, as we also noted in the previous subsections, this specific case is one for which the velocity of the Sedov-Taylor blastwave is proportional to the escape speed, and hence introducing a point mass to the problem does not violate self-similarity. In addition, the total energy of the system is still conserved.

Our results of Section \ref{sec:winds} agree with the findings of \citet{chevalier89}, who found that there is, in general, a contact discontinuity within the flow when $\gamma_1 = \gamma_2 \equiv \gamma \neq 1.5$, and that the flow extends through the contact discontinuity when $\gamma = 1.5$ (though it is not clear if the analogue with the Parker wind, and the existence of a sonic point within the flow, was also realized). \citet{chevalier89} also analyzed the case of $\gamma = 4/3$ and a density falloff of $\rho \propto r^{-2}$, and noted that for this case the solution extends to the origin and there is no contact discontinuity. We agree with this finding: as $\gamma \rightarrow 4/3$, the contact discontinuity approaches the origin, and disappears completely when $\gamma = 4/3$. The behavior arises, as noted by \citet{chevalier89}, from the fact that the binding energy of the ambient medium is zero when the density falls off as $\rho \propto r^{-2}$ and $\gamma = 4/3$ (note that, when our $\gamma_1 = \gamma_2 \neq 1+1/n$, the binding energy of the ambient medium on the right-hand side of Equation \ref{energytot} is $1/((n+1)(\gamma_2-1)) - 1$, which is zero when $n = 2$ and $\gamma_2 = 4/3$). In this case, then, there is no additional energy source needed at the origin, and there is no inner flow to give rise to the contact discontinuity. 

\citet{chevalier89} also analyzed the case of an outward moving accretion shock (his Section 4) that satisfies $r_{\rm sh} \propto t^{2/3}$, and looked for self-similar solutions for the velocity, density, and pressure behind the shock when the pre-shock material is in freefall. \citet{kazhdan94} -- the other paper with which we are familiar that looked for shock solutions that satisfy $r_{\rm sh} \propto t^{2/3}$ -- extended this work to include a range of boundary conditions (e.g., time-dependent flows without a shock). Both papers came to the conclusion that there was a critical shock propagation velocity that allowed the flow to settle into an equilibrium near the neutron star surface. \citet{kazhdan94} also considered alternative density profiles from $\rho \propto r^{-2}$, but only those that are shallower, and neither author seems to have obtained the solutions we presented here. 

In the next section we apply our self-similar shock solutions to \emph{failed} supernovae, where an outgoing sound wave only steepens into a shock near the base of the hydrogen envelope of a supergiant. In this problem the weakness of the outgoing shock is manifest and, as a consequence, the gravitational field of the central object and the gravitational potential energy of the ambient medium are both dominant contributors to the shock propagation, and the initial energy alone is not sufficient to determine the evolution of the post-shock fluid. 

\section{Application to failed supernovae and comparison to simulations}
\label{sec:simulations}
\subsection{Yellow supergiant}
Here we compare the predictions of our self-similar shock propagation model to the results of simulations of failed supernovae from \citet{fernandez18}. We initially focus on the 22 $M_{\odot}$ zero-age main sequence (ZAMS), yellow supergiant (YSG), which was evolved to core collapse with the {\sc mesa} stellar evolution code \citep{paxton11, paxton13, paxton15, paxton18}. The supernova then ``failed,'' which was mimicked by replacing the core of the star with a time-dependent, central mass, the mass itself evolving through the accretion of surrounding material and the (parameterized) emission of neutrinos until the core mass exceeded the Tolman-Oppenheimer-Volkov limit, at which point the neutrino emission ceased. The hydrodynamics of the response of the envelope to the mass loss, including the generation and propagation of a sound wave, the steepening of the wave into a shock, and the propagation of the shock through the overlying stellar material, was modeled with the {\sc flash} hydrodynamics code \citep{fryxell00}. For more details of the simulation itself, the prescription for the mass loss, etc., we refer the reader to \citet{fernandez18}. 

\begin{figure}[htbp] 
   \centering
   \includegraphics[width=0.495\textwidth]{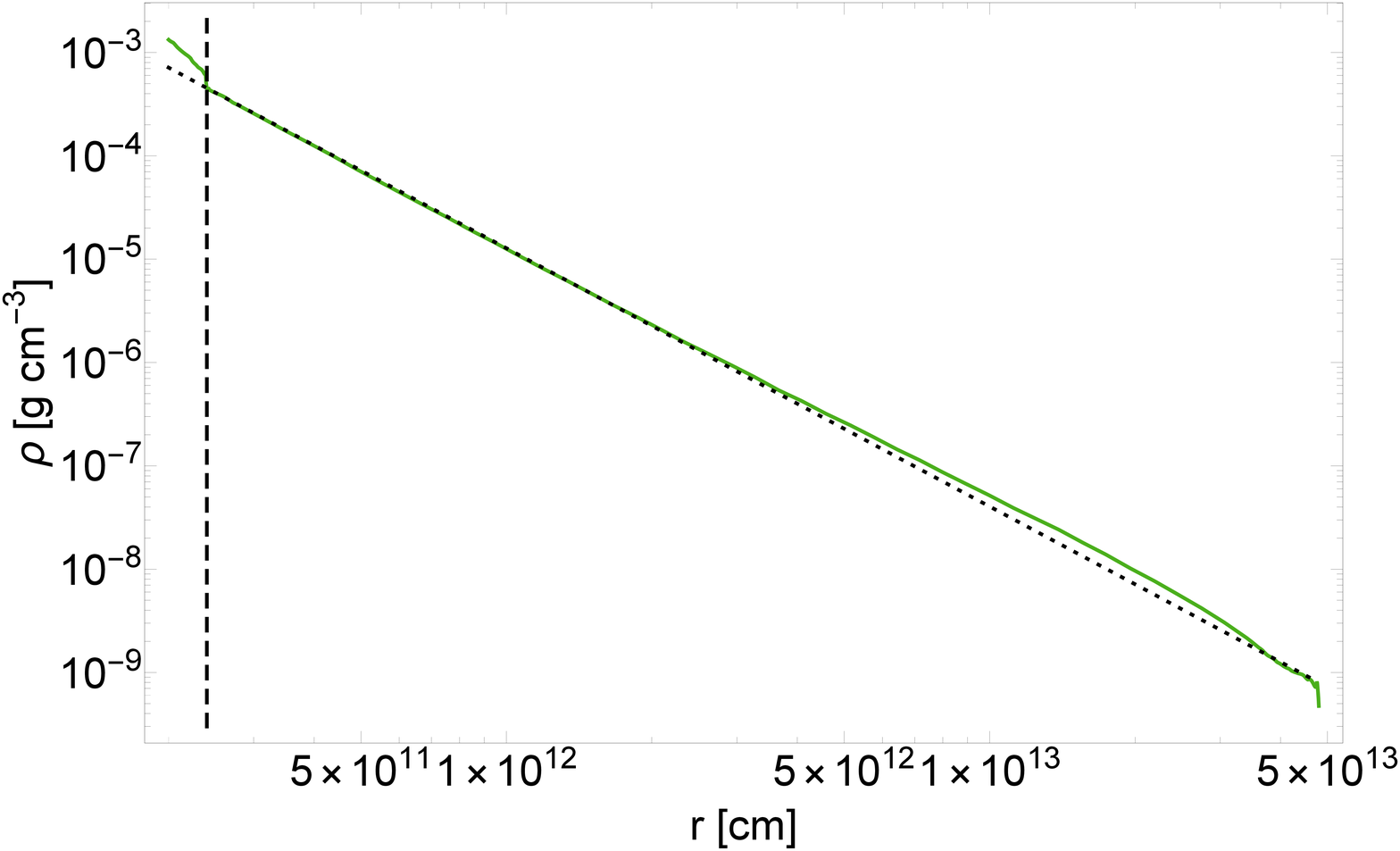} 
   \includegraphics[width=0.495\textwidth]{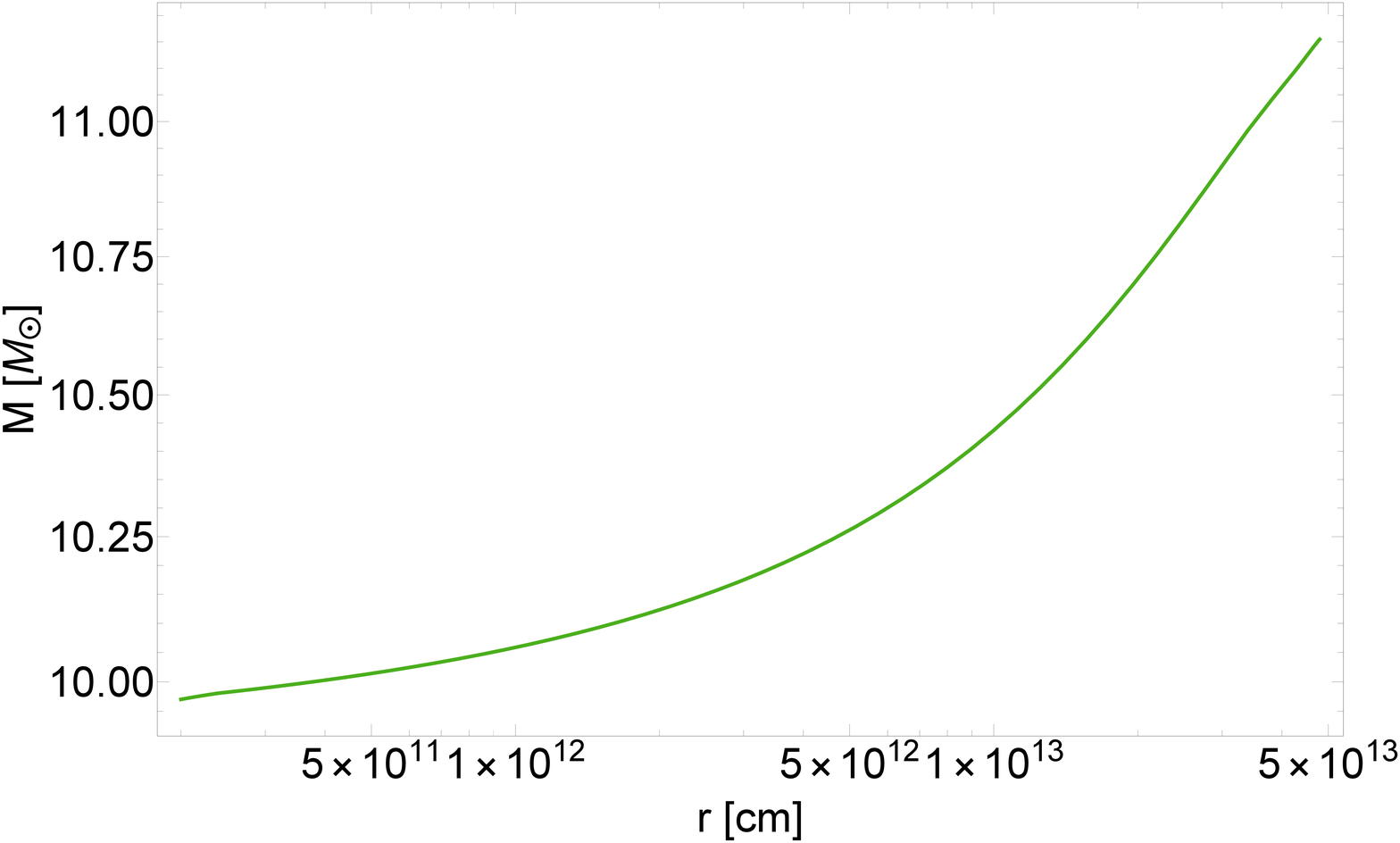} 
   \caption{Left: The density profile of a 22 $M_{\odot}$ ZAMS yellow supergiant at core-collapse (green curve) and the power-law $\rho \propto r^{-2.5}$ (dotted line); the vertical, dashed line shows the end of the helium core and the beginning of the hydrogen envelope. Right: The enclosed mass of the YSG as a function of radius. }
   \label{fig:ysgdata}
\end{figure}

The left-hand panel of Figure \ref{fig:ysgdata} shows the density profile of the YSG (green curve), and demonstrates that the falloff of the density within the hydrogen envelope, which begins at a radius of $r_0 \simeq 2.4\times 10^{11}$ cm (vertical, dashed line), is very well matched by the power law $\rho \propto r^{-2.5}$ (dotted line). The right-hand panel of this figure gives the mass profile of the YSG, which illustrates that the mass is nearly constant -- increasing only by 10\% over two orders of magnitude in radius -- once the hydrogen envelope is reached. The properties of this progenitor therefore agree very well with the underlying assumptions of our self-similar shock propagation model; though we have not plotted it, the adiabatic index $\gamma_1$ is also very close to $1.4 = 1+1/2.5$, so the adiabatic index of the gas is nearly equal to the polytropic index.

\begin{figure}[htbp] 
   \centering
   \includegraphics[width=0.495\textwidth]{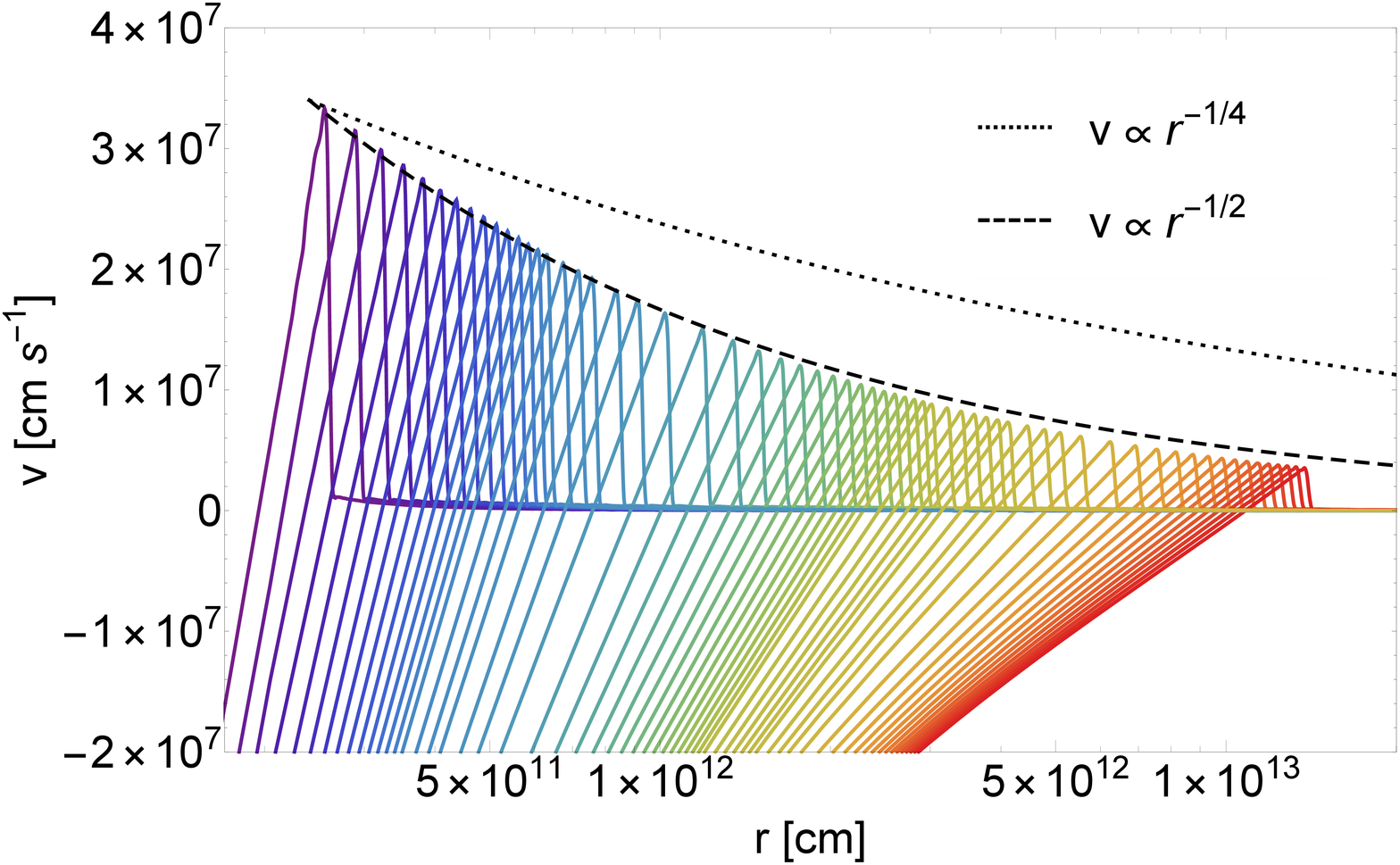} 
   \includegraphics[width=0.495\textwidth]{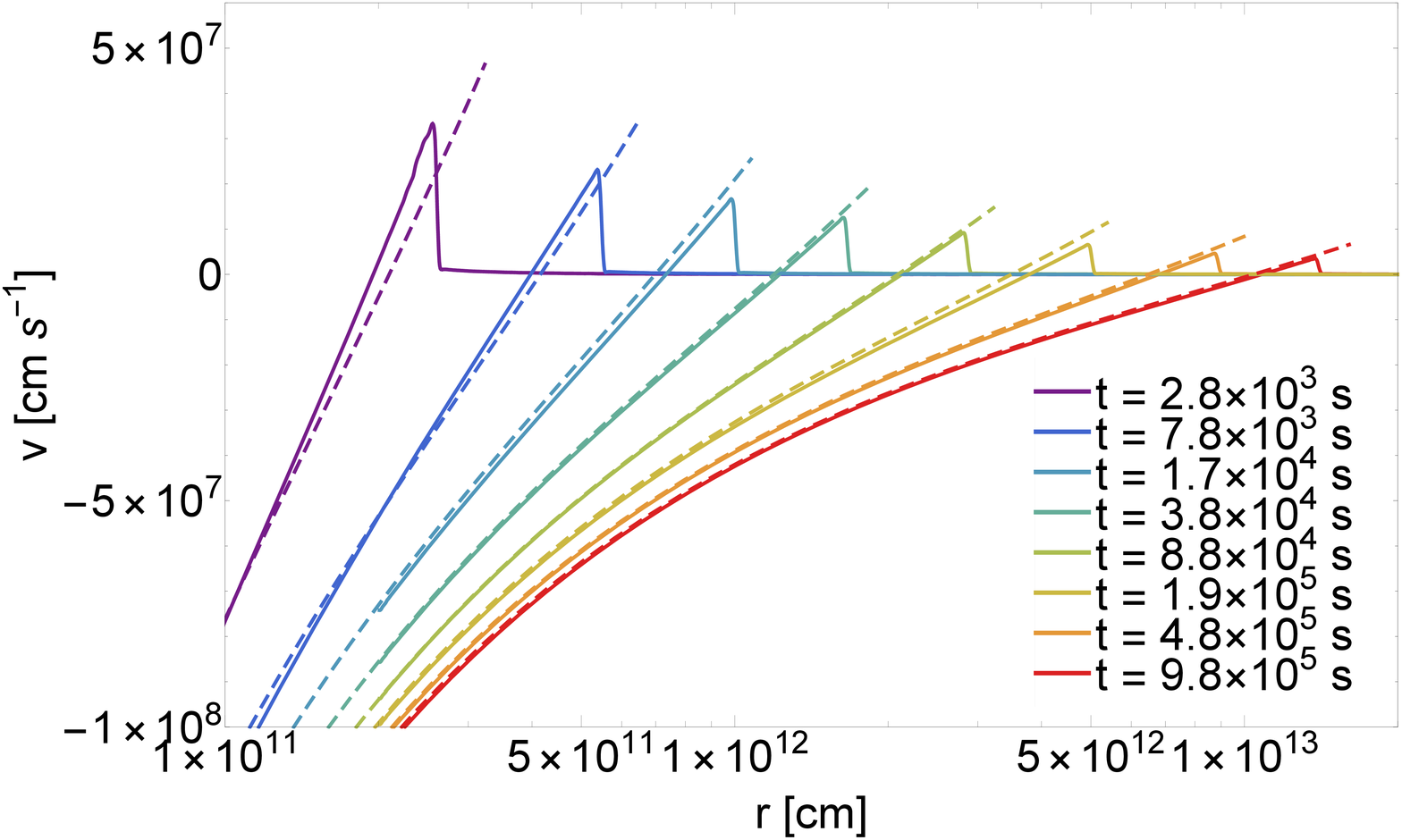}
   \caption{Left: The numerical solution for the velocity profile of the failed supernova from a 22 $M_{\odot}$ ZAMS yellow supergiant, analyzed numerically in \citet{fernandez18} (solid curves), which illustrates the propagation of a shock through the hydrogen envelope of the progenitor at a number of different times; the dashed curve shows the $v \propto r^{-1/2}$ scaling predicted from the self-similar solution, while the dotted curve shows the $r^{-1/5}$ solution that follows from the Sedov-Taylor blastwave, and each is normalized to match the velocity at the earliest time, which corresponds to the time at which the shock encounters the hydrogen envelope. Right: The numerical velocity profile at the times shown in the legend (solid curves), and the analytic prediction from the self-similar solution that passes through the sonic point with $n = 2.5$ and $\gamma = 1.4$ at the same times. }
   \label{fig:velocity}
\end{figure}

In the left-hand panel of Figure \ref{fig:velocity} we plot the numerical solution of the velocity from the simulation in \citet{fernandez18} for a finely-sampled set of points in time, the earliest corresponding roughly to the time at which the shock encounters the hydrogen envelope (left-most violet curve) and the latest when the shock is approaching the surface of the star (right-most red curve). The dotted line gives the prediction of the velocity scaling with radius that follows from the Sedov-Taylor (i.e., energy conserving) blastwave in a medium that satisfies $\rho \propto r^{-2.5}$, being $v\propto r^{-1/4}$, and the dashed curve gives the $v \propto r^{-1/2}$ scaling that results from our self-similar solutions. Each curve was normalized to match the velocity at the earliest time. We see not only that the Sedov-Taylor blastwave significantly overpredicts the magnitude of the velocity, but our self-similar solution very accurately reproduces the falloff of the shock speed as it traverses the hydrogen envelope, yielding discrepancies that amount to a percent at most (note that this figure is on a log-linear scale).

\begin{figure}[htbp] 
   \centering
   \includegraphics[width=0.495\textwidth]{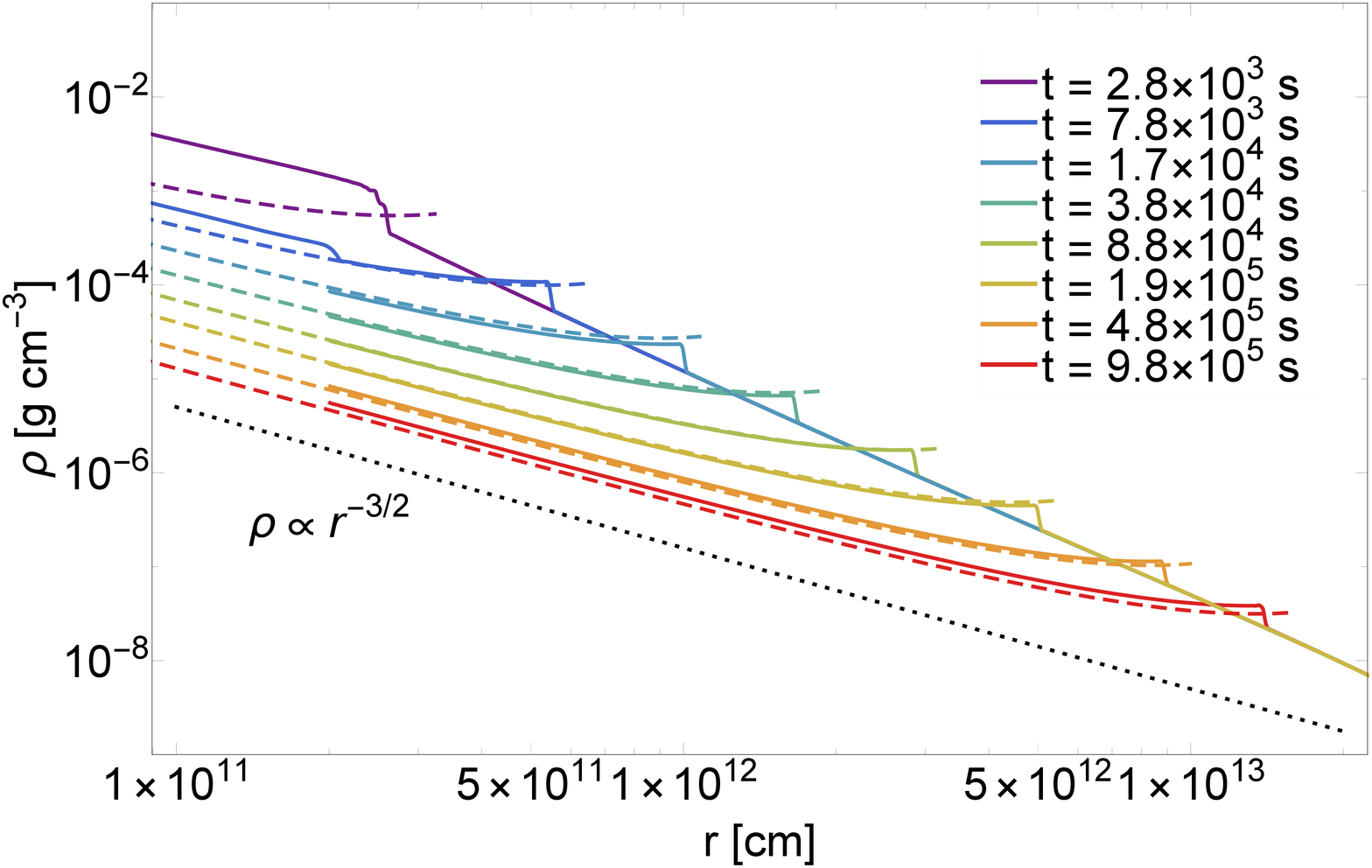} 
   \includegraphics[width=0.495\textwidth]{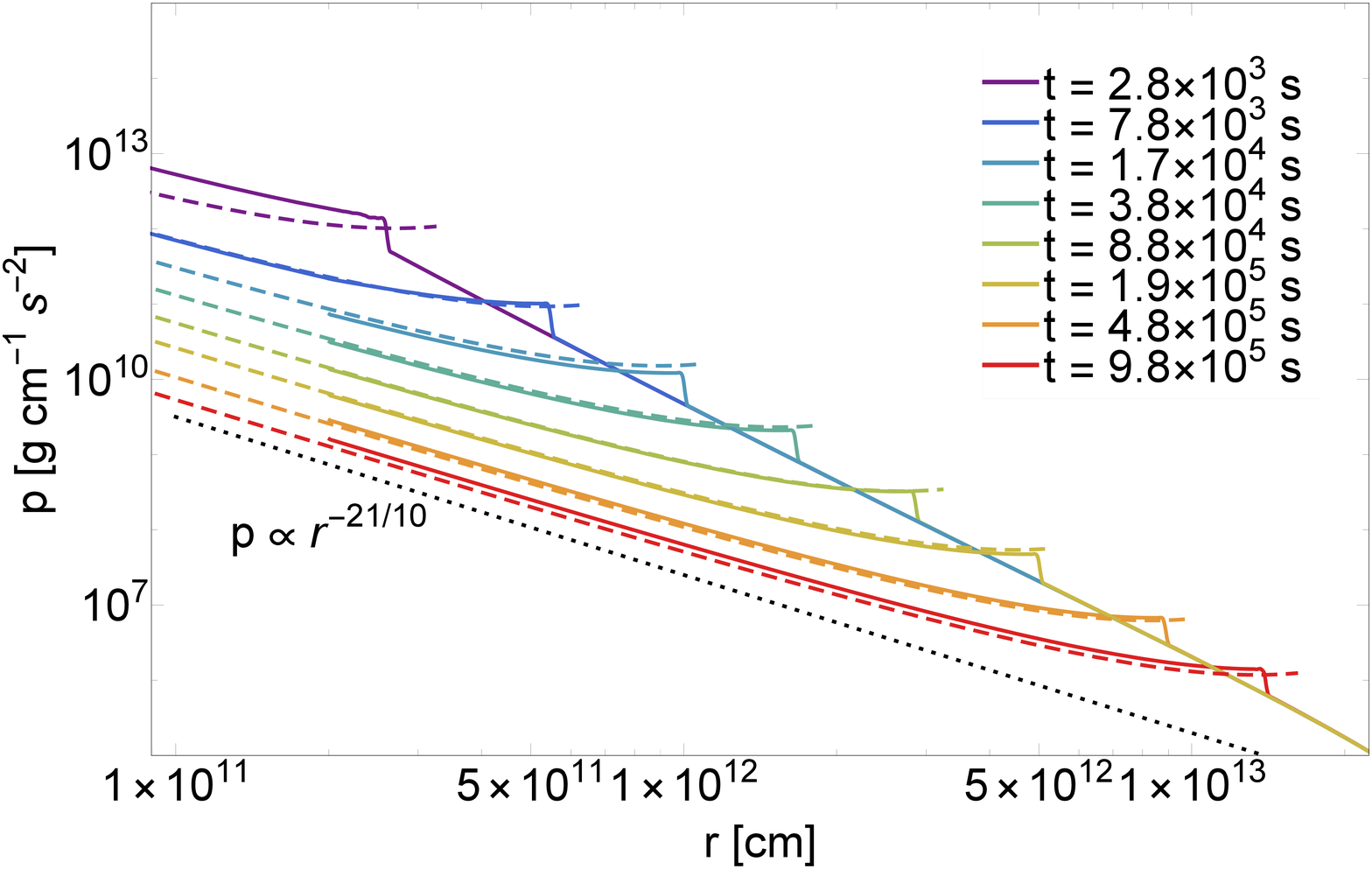}
   \caption{Left: The density profile from the numerical simulation of a failed supernova from a 22 $M_{\odot}$ ZAMS yellow supergiant, analyzed in \citet{fernandez18} (solid curves), and our self-similar, analytical prediction (dashed curves) with $n = 2.5$ and $\gamma = 1.4$ that passes through the sonic point; the different curves correspond to the times shown in the legend{}{, and the dotted line gives the asymptotic, $r\rightarrow 0$ scaling that results from freefall onto the black hole}. Right: The numerical solution for the pressure profile from the same simulation of the failed supernova as in the left-hand panel (solid curves), and the prediction from our self-similar model (dashed curves) for $n = 2.5$ and $\gamma = 1.4${}{; as for the density, the dotted curve gives the scaling of the pressure near the black hole}.}
   \label{fig:density}
\end{figure}

\begin{figure}[htbp] 
   \centering
   \includegraphics[width=0.995\textwidth]{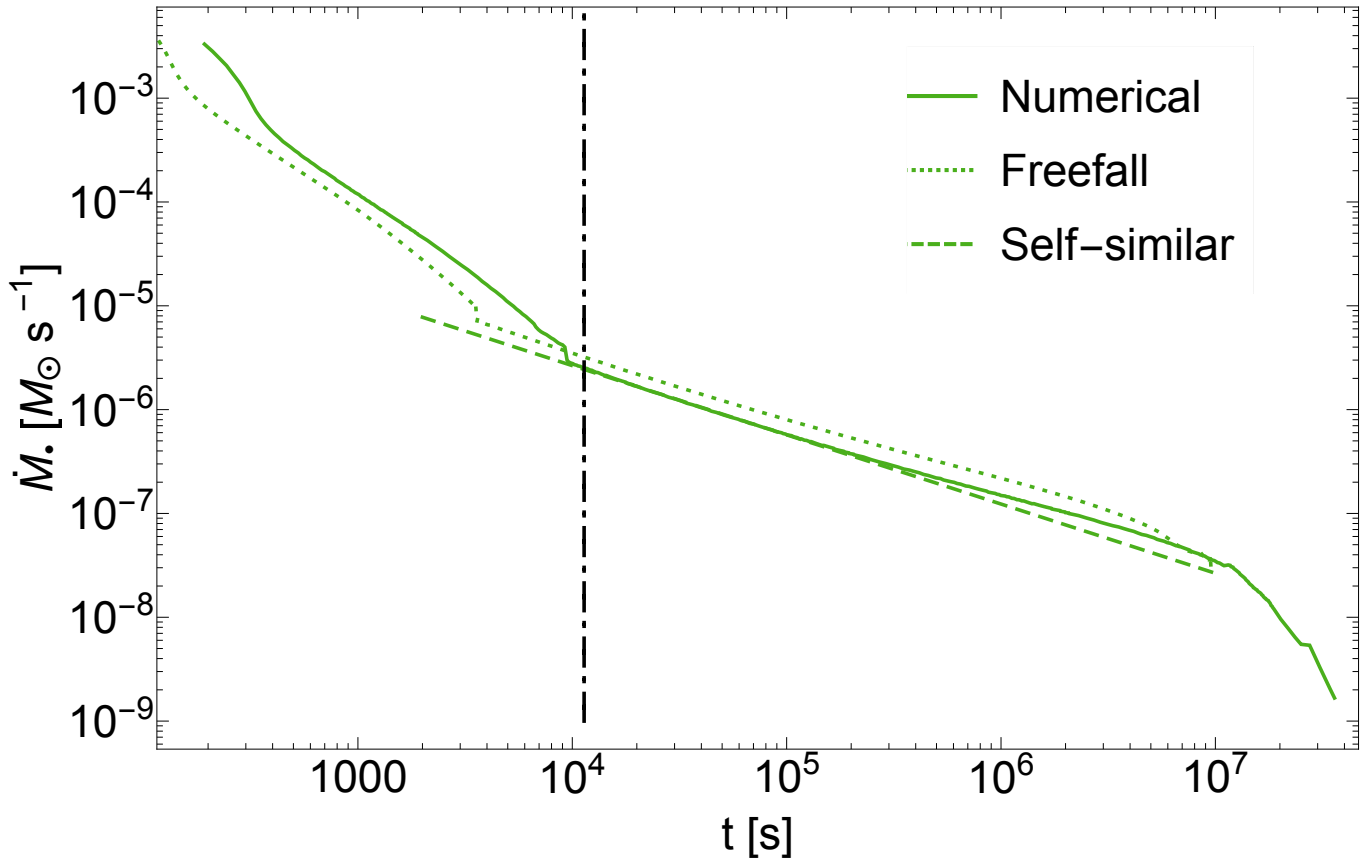} 
   \caption{Left: The accretion rate onto the black hole from the numerical simulation of the 22 $M_{\odot}$, ZAMS yellow supergiant presented in \citet{fernandez18} (solid curve), and the prediction from our self-similar solution (dashed curve), where the latter is given explicitly by Equation \eqref{mdot}, with $n = 2.5$, $\gamma = 1.4$ and $V = V_{\rm c} \simeq 1.202$, and scales as $\dot{M}_{\bullet} \propto t^{-2/3}$; the dotted line is the accretion rate if the entire progenitor freefalls from rest. The vertical, dashed line gives the predicted time at which this power-law phase should begin, being $t \simeq 1.1\times 10^4 $s, and is the result of solving Equation \eqref{lag} for the fallback time of the material immediately behind the shock when the shock is at the base of the hydrogen envelope of the YSG.}
   \label{fig:mdots}
\end{figure}

The right-hand panel of Figure \ref{fig:velocity} shows the numerical solution for the shock velocity (solid lines) at a smaller subset of times compared to the left-hand panel, the specific times shown in the legend. The dashed lines give the prediction of the self-similar model that passes through the sonic point and has $n = 2.5$ and $\gamma = 1.4$ ($= 1+1/n$), evaluated at the same times in the legend. Note that we are not renormalizing the self-similar velocity curve in any way, so there are no free parameters in this comparison. The agreement between the two, especially for later times, is excellent, with the only systematic and persistent difference being that the self-similar model slightly overpredicts the location of the shock front compared to the numerical solution. Some part of this discrepancy likely stems from the fact that the shock is established at the origin in the self-similar treatment, while in the numerical problem (and in any realistic, failed supernova) the shock is at a finite radius at $t = 0$; there will therefore be some difference between the two owing to initial conditions. The shock properties could also be slightly underresolved in the numerical simulation, as \citet{fernandez18} demonstrated that increasing the grid resolution in other simulations of failed supernovae had a non-negligible effect (at the 10\% level) on the shock strength and the fluid properties.

The left and right panels of Figure \ref{fig:density} show the density and pressure, respectively, from the numerical simulation (solid lines) and the self-similar model (dashed lines), again with $n = 2.5$ and $\gamma = 1.4$ and that pass through the sonic point, at the same times as in Figure \ref{fig:velocity}. For the self-similar model, we set $r_0 = 2.4\times10^{11}$ cm, the radius of the base of the hydrogen envelope of the YSG, and $\rho_0 = 4.79 \times 10^{-4}$ g cm$^{-3}$, the density of the YSG at $r_0$. As was true for the velocity, the agreement between the predictions of the self-similar model and the numerical simulation are excellent, the most noticeable difference being at late times when the self-similar model slightly underpredicts the numerical value. This could be due to small changes in the adiabatic indices, as gas pressure does become more important at lower temperatures and convection starts to become inefficient near the surface of the star, and/or the greater influence of self-gravity at these large radii.

The solid, green curve in Figure \ref{fig:mdots} gives the accretion rate, in solar masses per second, onto the black hole from the numerical simulation, strictly defined by the flux of mass through the radius $r = 2\times 10^{8}$ cm (this is actually a ``fallback rate,'' which could differ from the true accretion rate onto the black hole depending on the amount of rotation, disk microphysics, and magnetic fields, but we ignore these subtleties and use the two terms interchangeably). The dashed, green curve in this figure is the fallback rate predicted from the self-similar model, calculated from Equation \eqref{mdotbh} with $\gamma = 1.4$, $r_0 = 2.3\times 10^{11}$, and $\rho_0 = 4.79\times 10^{-4}$ g cm$^{-3}$ (again, the radius of the base of the hydrogen envelope and the density at that radius). In this case the limit appearing in Equation \eqref{mdotbh} is $\simeq 0.54$, and the accretion rate scales as $\dot{M}_{\bullet} \propto t^{-2/3}$. The dotted curve is the accretion rate if the entire star freefalls from rest under its own self-gravity, and can be obtained by making a suitable change of variables in the momentum and continuity equations; see \citet{coughlin17} for more details. The vertical, dot-dashed line gives our prediction for the time at which the power-law accretion rate should begin, which is determined by solving Equation \eqref{lag} for the fallback time of the fluid element immediately behind the shock when the shock is at the base of the hydrogen envelope. Numerically we find that this time corresponds to $\simeq 1.13\times 10^{4}$ s.

This figure demonstrates that the self-similar model reproduces both the normalization and the power-law decline of the numerical fallback rate almost exactly, with visible, but small, deviations arising only at times $\gtrsim few \times 10^{5}$ s. These differences are almost certainly due to the change in the power-law index of the density profile at larger radii, with the shallower falloff -- see Figure \ref{fig:ysgdata} -- yielding a slightly higher accretion rate. The time at which the $t^{-2/3}$ decline begins is also very well reproduced by calculating the fallback time of the mass shell at the base of the hydrogen envelope using the self-similar velocity profile. The small difference between the self-similar prediction and the numerical finding is likely due to the fact that the shock reaches the base of the hydrogen envelope in the numerical simulation at a slightly different time than that prescribed by the self-similar model. The freefall solution also reproduces the power-law decline of the accretion rate, and, as we noted in Section \ref{sec:scalings}, this is due to the fact that the velocity and density profiles in the self-similar model approach their freefall scalings near the point mass. However, the freefall rate does not reproduce the correct normalization and underpredicts the time at which the power-law begins by an order of magnitude; this discrepancy arises because the freefall solution assumes that the fluid everywhere starts to collapse simultaneously at $t = 0$, whereas, in actuality, the information about the collapsing core is communicated by the shock on roughly the sound crossing time (but slightly shorter owing to the supersonic propagation of the shock). 

The collection of Figures \ref{fig:velocity} -- \ref{fig:mdots} illustrate that the fluid dynamics following the failed supernova of a YSG, analyzed numerically in \citet{fernandez18}, are very well reproduced by our self-similar shock propagation model, both at very different radii and times. This agreement strongly suggests that the self-similarity of the flow behind the shock is established in these systems on timescales shortly after the shock penetrates into the hydrogen envelope of the progenitor. 

\begin{figure}[htbp] 
   \centering
   \includegraphics[width=\textwidth]{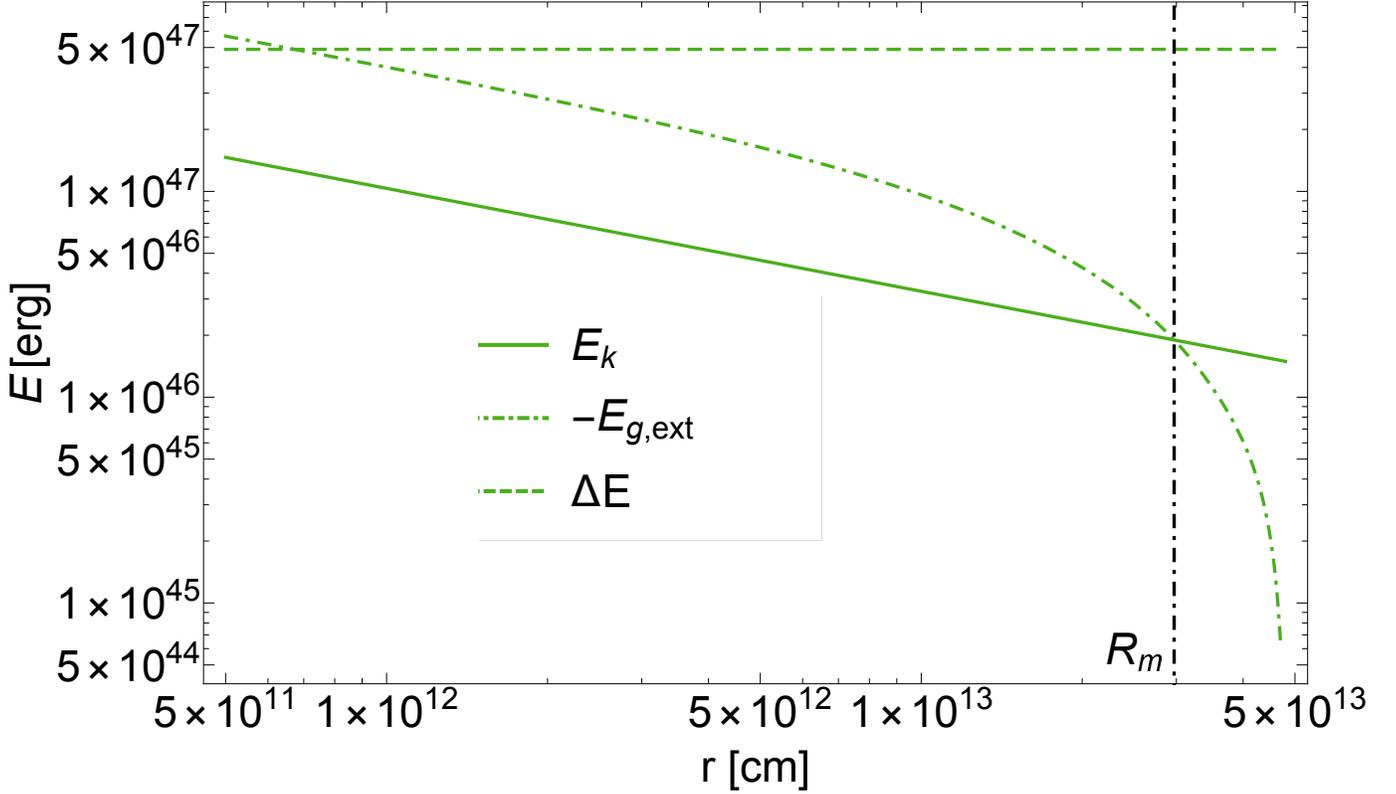} 
   \caption{{}{The dot-dashed curve shows the gravitational binding energy exterior to radius $r$ (see Equation \ref{Egext}) for the 22 $M_{\odot}$, ZAMS YSG analyzed numerically in \citet{fernandez18}, and the solid curve gives the self-similar, kinetic energy contained in the self-similar blast wave when $\gamma = 1.4$ and $n = 2.5$, where $\rho \propto r^{-n}$. The radius at which these two are equal, $R_{\rm m} \simeq 3.0\times 10^{13}$ cm, denotes the location in the progenitor at which the swept up binding energy can no longer balance the kinetic energy of the blast, and therefore corresponds to the radius at which post-shock material will start to be kinetically-dominated and unbound from the point mass. The horizontal, dashed line is the energy liberated from the mass lost to neutrinos, analytically estimated to be $\Delta E \simeq 4.9\times 10^{47}$ erg \citep{fernandez18, coughlin18}, and is shown here to demonstrate that this energy overestimates the kinetic energy of the blast owing to the relative importance of the gravitational potential energy.}}
   \label{fig:extgrav}
\end{figure}

{}{Even though the 22 $M_{\odot}$ ZAMS YSG presented here and analyzed in \citet{fernandez18} yielded a questionable outcome in terms of the amount of mass ejected and the final energetics, we note that our self-similar solutions \emph{should} result in mass ejection: because any realistic star has a photosphere and a finite binding energy, there will always be some radius, $R_{\rm m}$, near the stellar surface where the binding energy exterior to the shock drops below the kinetic energy of the shocked fluid. Beyond this radius, the precise balance between the swept up gravitational energy and the kinetic energy behind the shock -- which is required by the self-similar solution -- cannot be maintained, and thereafter the shock will become kinetically dominated and produce a shell of unbound gas. Figure \ref{fig:extgrav} shows the magnitude of the gravitational binding energy of this progenitor exterior to $r$ (dot-dashed curve), }

\begin{equation}
E_{\rm g, ext} = -\int_{r}^{R_*}\frac{GM}{r} 4\pi r^2\rho\, dr \label{Egext}
\end{equation}
{}{and the self-similar kinetic energy behind the shock for $n = 2.5$ (solid curve; see also Equation \ref{Etot}), }

\begin{equation}
E_{\rm k} = 4\pi \rho_0 r_0^3\frac{GM}{r_0}\left(\frac{r}{r_0}\right)^{-\frac{1}{2}}\int_0^{1}\frac{1}{2}V^2\xi^{-3/2}gf^2 d\xi \simeq 0.486 \times 4\pi \rho_0 r_0^3\frac{GM}{r_0}\left(\frac{r}{r_0}\right)^{-\frac{1}{2}}
\end{equation}
{}{both as functions of $r$, for the 22 $M_{\odot}$ ZAMS YSG. The horizontal, dashed line is the initial energy liberated from the mass lost to neutrinos, $\Delta E \simeq 4.9\times 10^{47}$ erg, calculated analytically for this progenitor \citep{fernandez18, coughlin18}. }

{}{The radius at which these two are equal, $R_{\rm m} \simeq 3.0 \times 10^{13}$ cm, is shown by the vertical, dashed line in Figure \ref{fig:extgrav}. 
Note that this $R_{\rm m}$ differs substantially from the one that would be inferred by equating the \emph{initial} energy lost to neutrinos to the external gravitational potential energy, which, from this figure, is $R_{\rm m} \simeq 6\times 10^{11}$ cm -- nearly two orders of magnitude smaller than the one found by using the kinetic energy of the self-similar solution. This discrepancy highlights the fact that the gravitational potential energy of the star dramatically reduces the energy budget of the shock by the time the shock approaches the stellar surface. }

{}{We can also calculate $\Delta M$ -- the amount of mass exterior to $R_{\rm m}$ -- for this progenitor, which gives $\Delta M \simeq 0.24 M_{\odot}$. Since this material has positive kinetic energy following the passage of the shock, we expect this material to be successfully ejected and unbound from the black hole. While \citet{fernandez18} did not quote (in their Table 2) an estimate for the amount of ejected gas owing to uncertainties in the equation of state of the gas, they did note that ``\ldots{}the model ejects a small amount of mass $\sim 10^{-1} M_{\odot}$\ldots,'' which is in agreement with our predictions here. On the other hand, using the initial energy lost to neutrinos to calculate the total ejected mass gives $\Delta M \simeq 1.1 M_{\odot}$, which overestimates the amount of mass capable of being ejected from the star by an order of magnitude. }

{}{Finally, even though material will start to be unbound beyond the radius $R_{\rm m}$, there will still be some gas with a very small, but marginally negative, energy initially ejected from a radius around $R_{\rm m}$. This ``marginally-bound material'' will continue to fall back onto the black hole at late times, but with an accretion rate that scales as $\dot{M}_{\bullet} \propto t^{-5/3}$. Using the fact that gas parcels just interior to $R_{\rm m}$ should still approximately obey the self-similar solution, we can calculate from Equation \eqref{lag} the fallback time from $R_{\rm m}$, which we will denote $T_{\rm m}$, which should correspond to the time at which the accretion rate transitions from $t^{-2/3}$ to $t^{-5/3}$. Doing so gives $T_{\rm m} \simeq 1.8\times 10^{7}$ s; in Figure \ref{fig:mdots}, at times very close to $T_{\rm m}$ the numerically-obtained fallback rate deviates from the power-law decline and starts to fall off more rapidly with time.}

\subsection{Red supergiant}
\begin{figure}[htbp] 
   \centering
   \includegraphics[width=0.495\textwidth]{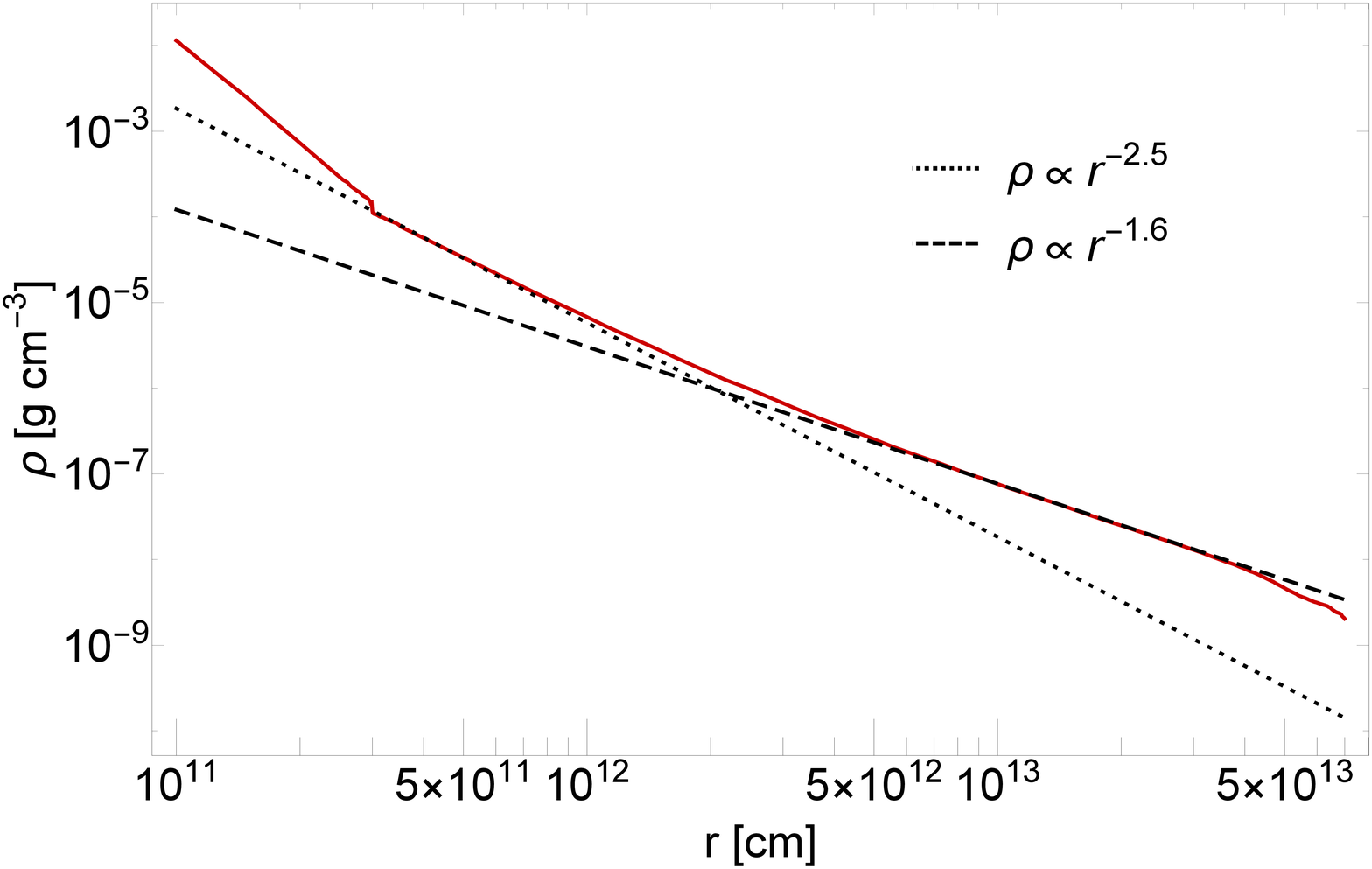} 
   \includegraphics[width=0.485\textwidth]{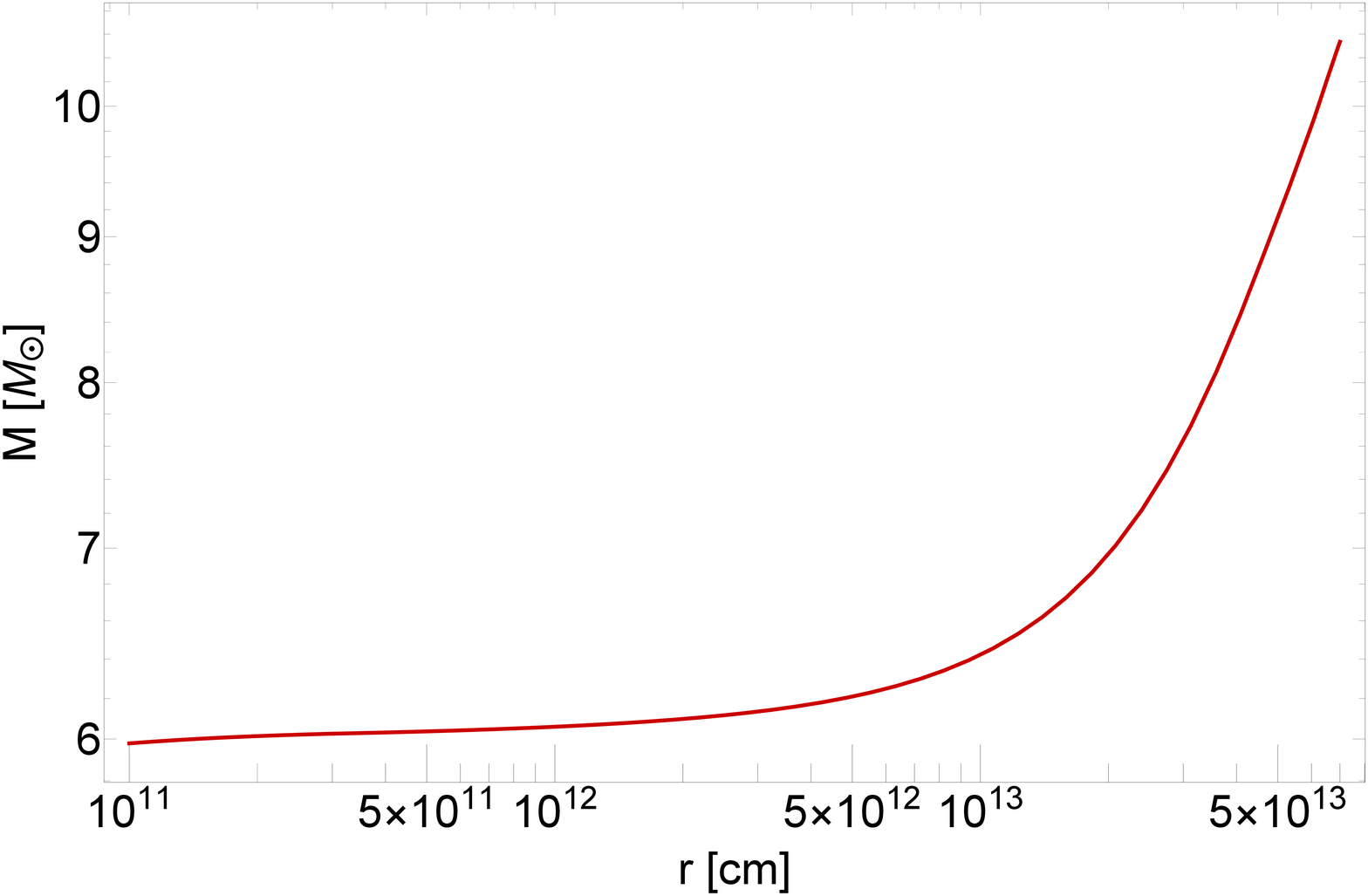} 
   \caption{The left panel shows the density profile at core collapse of the 15 $M_{\odot}$, ZAMS RSG analyzed in \citet{fernandez18}, and the right-hand panel gives the mass contained within $r$ for the same progenitor.}
   \label{fig:rsg_data}
\end{figure}

The YSG was one of the supergiants analyzed by \citet{fernandez18} that gave rise to a very weak shock (in terms of shock velocity upon encountering the hydrogen envelope), and consequently a very small amount of mass ejection (if any). There were other cases, however, that were not so ambiguous, one such case being the 15 $M_{\odot}$ ZAMS red supergiant (RSG), which ejected a large amount of mass ($\sim$ few $M_{\odot}$). The left-hand panel of Figure \ref{fig:rsg_data} shows the density profile of this star at the time of core collapse. The hydrogen envelope is well-described by a density profile that scales as $\rho \propto r^{-2.5}$ for radii not much larger than $r \simeq 3\times 10^{11}$ cm, which corresponds to the base of the hydrogen envelope. However, at a radius of roughly $10^{12}$ cm, the power-law index becomes shallower, and the falloff is better described by $\rho \propto r^{-1.6}$ for radii in the range $3\times 10^{12} \lesssim r \lesssim 4\times 10^{13}$ (even though the adiabatic index of the gas is still $\gamma_1 \simeq 1.4$). In the right-hand panel of this figure we plot the mass contained within $r$ of the RSG. In agreement with the fact that the density falls off less steeply, we see that the mass contained at large radii increases substantially for this progenitor.

\begin{figure}[htbp] 
   \centering
   \includegraphics[width=0.995\textwidth]{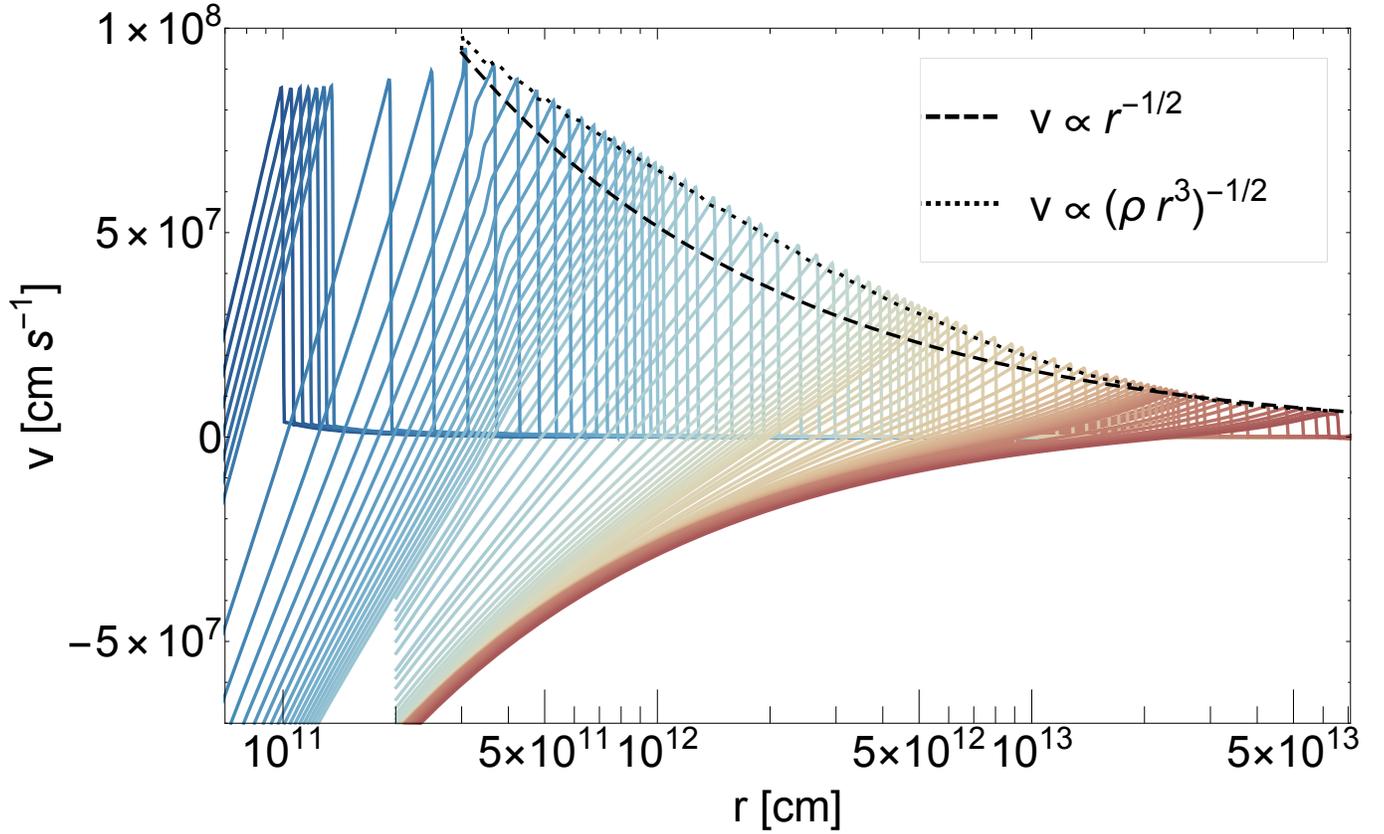}  
   \caption{The velocity as a function of $r$ over a finely-sampled range of times, given by the different curves, of the failed supernova from the 15 $M_{\odot}$, ZAMS red supergiant analyzed in \citet{fernandez18}; the dashed curve shows the $r^{-1/2}$ velocity scaling expected from our self-similar model, and the dotted curve shows the Sedov-Taylor velocity from Equation \eqref{stvel} with $\rho$ given by the density of the progenitor, and each is normalized to the peak in the velocity at the time the shock reaches the base of the hydrogen envelope at $r = 3\times 10^{11}$ cm;  the dotted curve follows $v \propto r^{-1/4}$ near the base of the hydrogen envelope where $\rho \propto r^{-2.5}$, while farther out -- where $\rho \propto r^{-1.6}$ -- it follows $v \propto r^{-0.7}$ (see Figure \ref{fig:rsg_data}). For this progenitor, the energy-conserving, Sedov-Taylor solution better matches the numerical results.} 
   \label{fig:vofr_rsg}
\end{figure}

Because this star does not satisfy the two basic assumptions of our self-similar prescription -- that the density is well-described by a single power law and that the enclosed mass is nearly constant -- we do not expect the numerical simulations of this failed supernova to agree as well with the self-similar model. Nevertheless, this case is still interesting, as the gravitational mass radiated in neutrinos for this model was roughly three times that of the YSG, corresponding to an increase in the shock energy (see Table 1 of \citealt{fernandez18}). The greater amount of mass at large radii in this star also implies that the gravitational potential is less significant near the base of the hydrogen envelope as compared to the YSG, as the two stars had roughly the same mass at the onset of core collapse. These two points imply that the shock propagation could, at least initially, be better described by the Sedov-Taylor blastwave. 

Figure \ref{fig:vofr_rsg} shows the velocity of the numerical simulation from \citet{fernandez18} over a finely-sampled range of times, given by the different curves; colors simply serve to distinguish the curves from one another. The dashed line gives the $v_{\rm sh} \propto r^{-1/2}$ scaling of our self-similar model, while the dotted line is $v_{\rm sh} \propto (\rho \, r^3)^{-1/2}$, which follows from the Sedov-Taylor blastwave (see Equation \ref{stvel}) with $\rho$ given by the density profile of the RSG, and each is normalized to the velocity of the shock as it encounters the hydrogen envelope at $r = 3\times 10^{11}$ cm. We see that the Sedov-Taylor scaling provides a very good fit to the falloff of the shock velocity as it propagates through the RSG, and gives $v_{\rm sh} \propto r^{-1/4}$ near the base of the hydrogen envelope (where the density declines approximately as $\rho \propto r^{-2.5}$) and $v_{\rm sh} \propto r^{-0.7}$ at larger radii where $\rho \propto r^{-1.6}$. 

\begin{figure}[htbp] 
   \centering
   \includegraphics[width=0.495\textwidth]{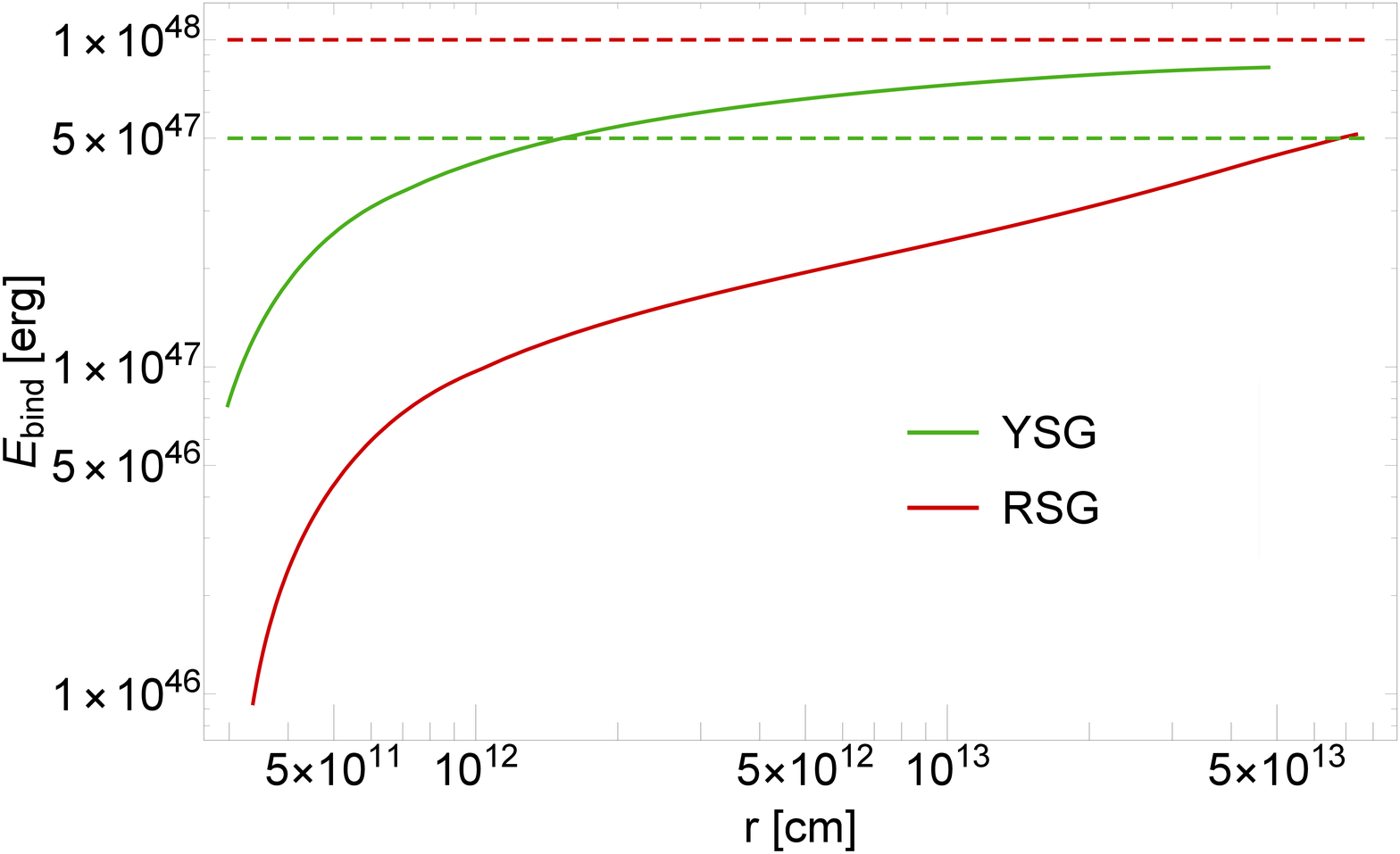} 
   \includegraphics[width=0.495\textwidth]{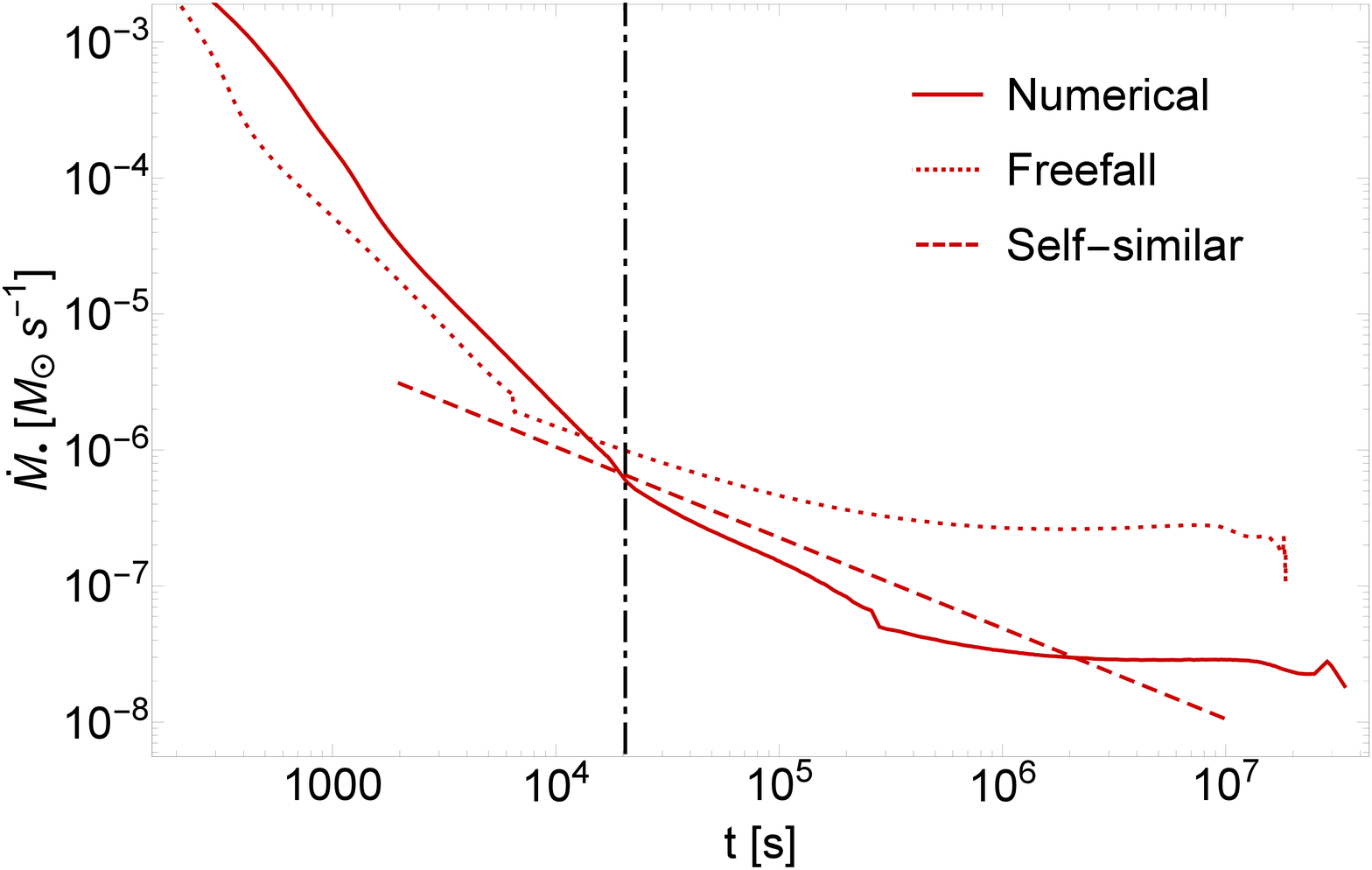} 
   \caption{Left: The total, integrated binding energy from the base of the hydrogen envelope of the 22 $M_{\odot}$ ZAMS yellow supergiant (green curve) and the 15 $M_{\odot}$ ZAMS red supergiant (red curve), the failed supernovae of which were both analyzed numerically in \citet{fernandez18}; the horizontal, dashed lines give the analytic predictions for the total energy contained in the shocked fluid, and demonstrate that the total shock energy is much greater than the binding energy for the RSG for the majority of the envelope. For this reason, we expect the shock to be in the Sedov-Taylor regime for the red supergiant, while the rough equality between the shock energy and the binding energy of the yellow supergiant implies that our self-similar solution is more applicable; we find, indeed, that this is the case. Right: The solid curve gives the accretion rate onto the black hole following the failed supernova of the $15 M_{\odot}$, ZAMS red supergiant, numerically obtained from the simulations in \citet{fernandez18}; the dotted curve assumes that the entire star freefalls from rest, and the dashed curve is from our self-similar model with $n = 2.5$, $\gamma_1 = \gamma_2 = 1.4$, $r_0 = 3\times 10^{11}$ (corresponding to the radius of the base of the hydrogen envelope) and $\rho_0$ is the density of the progenitor at $r_0$. The vertical, dot-dashed line gives the prediction of the time at which we expect the power-law fallback to begin, corresponding to the time taken for the fluid element immediately behind the shock to reach the origin when the shock reaches the base of the hydrogen envelope.}
   \label{fig:bind}
\end{figure}

The left-hand panel of Figure \ref{fig:bind} gives the binding energy of the material exterior to the base of the hydrogen envelope, being

\begin{equation}
E_{\rm bind} = \int_{r_0}^{r}\frac{GM(r)}{r}4\pi r^2\rho dr,
\end{equation}
where $r_0 = 2.4\times 10^{11}$ cm for the YSG (green curve in this figure) and $r_0 = 3\times 10^{11}$ cm for the RSG (red curve). The dashed lines give the analytic predictions for the total energy contained behind the shock, which scales as the square of the mass lost to neutrinos (it also depends less sensitively on the stellar structure; see \citet{coughlin18} and column 11 of Table 2 of \citet{fernandez18} for more details). This demonstrates, as we argued above, that the YSG has a larger binding energy near the base of its hydrogen envelope due to the fact that the mass interior to $r_0$ is larger. For this reason and because the neutrino mass loss energy is less in this case, the shock does not experience any phase in which the Sedov-Taylor phase describes the shock propagation. For the RSG, on the other hand, the shock experiences less of a gravitational barrier upon reaching the hydrogen envelope, and this feature coupled to the larger amount of intrinsic energy from the neutrino mass loss causes the shock to follow the Sedov-Taylor solution.

The right-hand panel of Figure \ref{fig:bind} gives the accretion rate onto the black hole following the failed supernova of the RSG, where the solid curve is from the simulation of \citet{fernandez18}, the dotted curve assumes that the star freefalls from rest, and the dashed curve is from the self-similar solution with $n = 2.5$ and $\gamma = 1.4$. As for the self-similar accretion rate that we calculated from the YSG, in Equation \eqref{mdot} we used $r_0 = 3\times 10^{11}$ cm, being the radius corresponding to the base of the hydrogen envelope, and $\rho_0$ is the density of the progenitor at that radius. The vertical, dot-dashed line in this figure is the prediction for the time at which the power-law falloff should begin, which follows from the solution to Equation \eqref{lag} and is the time taken for the fluid parcel immediately behind the shock front to reach the origin when the shock encounters the hydrogen envelope. We see that until a time of $\sim 3\times 10^5$ s, the self-similar solution slightly overpredicts the normalization of the fallback rate, and the accretion rate from the numerical simulation falls off slightly steeper than the predicted one of $\dot{M}_{\bullet} \propto t^{-2/3}$. After this time the numerical fallback rate transitions to a much flatter profile, which likely corresponds to the point at which the shock encounters the region of the envelope better characterized by $\rho \propto r^{-1.6}$. This hypothesis is substantiated by the fact that the freefall accretion rate should be flat when $n = 5/3$, and this is reflected by the freefall accretion curve. 

The agreement between the self-similar fallback model and the numerical solution at early times is somewhat surprising: from Figure \ref{fig:vofr_rsg}, the shock velocity is well-described by the Sedov-Taylor scaling, which implies that the kinetic energy of the material behind the shock is larger than the gravitational potential energy. We might therefore suspect that material falling back to the black hole is only very weakly bound, and should thus satisfy $\dot{M}_{\bullet} \propto t^{-5/3}$; indeed, this asymptotic scaling is upheld by the failed supernova simulations of blue supergiants (BSGs) and Wolf-Rayets (WRs), also performed by \citet{fernandez18}, which possess only a small hydrogen envelope (BSG) or lack one entirely (WR). In these cases, the shock accelerates rapidly down the density gradient of the star, erupts from the surface and ejects material from the system, whereafter it adiabatically expands and cools to the point where the kinetic energy dominates the internal energy; see Figures 4 and 10 of \citet{fernandez18}.

The fact that the fallback rate for the RSG does not follow $t^{-5/3}$ implies that, despite the fact that the kinetic energy outweighs the gravitational potential energy (and hence the total energy is roughly conserved and the shock propagates in the Sedov-Taylor regime), the apocenter distances of the fluid parcels are not significantly larger than their initial radii within the progenitor. Put another way, even though the velocity profile must fall into the Sedov-Taylor regime near the shock front, there is a finite distance behind the shock to which the fluid elements fall where the gravitational potential energy is not ignorable, and the fluid is still gravitationally bound to the black hole (see the post-shock velocity profiles in Figure \ref{fig:vofr_rsg}). For this reason, the evolution of the shocked fluid is probably not completely self-similar in this case, and the radius interior to which the gravitational field contributes significantly to the energetics evolves in time in a non-self-similar way. 

This finding also implies that if the initial shock energy had been smaller (or if the envelope of the RSG extended to yet larger radii), then it is likely that the shock would not have been able to maintain the Sedov-Taylor phase throughout the entire hydrogen envelope, and the binding energy swept up from the ambient medium would have eventually dominated the energetics of the flow; this conclusion is supported by Figure 5 of \citet{fernandez18}, which shows that the total energy budget of the shocked material is roughly conserved over the majority of the shock propagation, but does start to decline prior to reaching the stellar photosphere.  As we noted in Section \ref{sec:scalings}, such a shallow density profile and a small initial kinetic energy likely results in the dissolution of the shock and no mass ejection. We suspect that this outcome likely occurred for the 25 $M_{\odot}$ ZAMS YSG and the 80 $M_{\odot}$ ZAMS BSG, the failed supernovae of which were also investigated numerically in \citet{fernandez18}, which completely failed and did not eject any mass with positive energy.

\section{Summary and Conclusions}
\label{sec:conclusions}
In this paper we described a model for the self-similar propagation of a mild to intermediate strength shock (i.e., one with a Mach number of order a few) through a power-law ambient medium that is both adiabatic and in hydrostatic equilibrium. In contrast to the Sedov-Taylor blastwave, these solutions account both for the gravitational field of a central object and the gravitational binding energy of the surrounding gas, neither of which is ignorable when the Mach number of the shock is not large. 

We found that, if the power-law falloff of the ambient medium is steeper than $\rho \propto r^{-2}$, then there are self-similar solutions for the fluid variables (the velocity, density, and pressure) behind an outward-propagating shock that extend to the origin and yield accretion onto the central object. These solutions require a specific value of the shock velocity that permits the smooth passage of the flow through a sonic point. In analogy with the Bondi problem, solutions with shock velocities smaller than this critical one ``settle'' into hydrostatic balance near the origin and are likely unachievable when the point mass is a black hole (owing to the diverging pressure gradient near the origin), while those with larger velocities are double-valued and are physically impossible. When the power-law index of the density profile of the ambient medium falls below 2, there are no longer solutions that yield accretion onto the point mass owing to the infinite amount of gravitational potential energy of the ambient medium; when the power-law index is exactly 2, the only solution that does not contain an energy source is the Sedov-Taylor blastwave. There are, however, solutions in this latter regime that again pass through a critical point and have outflow near the point mass, and therefore include a source of energy at the origin and are analogues of the Parker wind. The energy input from the central source must exactly scale as the influx of gravitational binding energy from the ambient medium in these cases, and are therefore unlikely to be realized in nature where the energy generation rate is dictated by local microphysics near the central object.

We applied our model to a failed supernova -- when core collapse fails to unbind the overlying stellar envelope, but a ``sound pulse,'' generated from neutrino-induced mass loss, steepens into a weak shock in the outer envelope of the star. We focused on the 22 $M_{\odot}$ (ZAMS) yellow supergiant (YSG) progenitor analyzed numerically in \citet{fernandez18}, the hydrogen envelope of which is very well-matched by a polytrope with polytropic index $\gamma = 1.4$, and is therefore characterized by $\rho \propto r^{-2.5}$ and $p \propto r^{-3.5}$ (see Figure \ref{fig:ysgdata}). Our predictions for the overall scaling of the shock velocity, the time and space-dependent variations of the post-shock velocity, density, and pressure, and the accretion rate onto the black hole agree very well with the simulation data; see Figures \ref{fig:velocity} -- \ref{fig:mdots}. This agreement on all fronts almost certainly indicates that this mode of self-similar shock propagation occurs in this failed supernova. 

We also investigated the failed supernova from the 15 $M_{\odot}$ ZAMS red supergiant (RSG) presented in \citet{fernandez18}. Unlike the YSG, the density profile of this star is not as well modeled by a single power-law, and instead flattens from $\rho \propto r^{-2.5}$ to $\rho \propto r^{-1.6}$ at larger radii within the hydrogen envelope (see Figure \ref{fig:rsg_data}). The amount of energy injected into the initial sound pulse was also greater for this case and the gravitational binding energy at the base of the hydrogen envelope was smaller, as shown in the left-hand panel of Figure \ref{fig:bind}; these two points together caused the shock to initially propagate in the Sedov-Taylor phase, which is substantiated by Figure \ref{fig:rsg_data}. However, the fallback onto the black hole (right-hand panel of Figure \ref{fig:bind}) is still fairly well matched by our self-similar prediction, and the accretion rate does not transition to $\dot{M}_{\bullet} \propto t^{-5/3}$ -- the power-law fallback rate that would result if the material were only weakly bound to the black hole. This finding suggests that, while the overall energy in this case was large enough to allow the shock velocity to be better described by the Sedov-Taylor blastwave, the flow is not yet unbound from the system and the entire, post-shock fluid is likely not able to be described by a single self-similar variable. 

While the Sedov-Taylor blastwave uses the (conserved) energy of the explosion to determine the shock velocity, the self-similar solutions presented here require the energy to maintain a very specific, time-dependent value (see Equation \ref{Etot}). As we argued in Section \ref{sec:scalings}, when the shock energy is slightly above or slightly below this value as it encounters the hydrogen envelope, it is likely that the system can self-regulate -- through accretion and sweeping up the binding energy of the ambient medium -- to the point where the self-similar energy is achieved {}{(though this argument is only a heuristic one; we plan to investigate the validity of this notion from a rigorous perturbation analysis in a future paper)}. However, if the initial shock energy is much greater than the binding energy of the ambient medium, then the energy of the shock cannot be reduced to arbitrarily-small values, and the shock will asymptotically follow the Sedov-Taylor solution as the kinetic energy dominates the potential energy. Similarly, if removing all of the bound material through accretion cannot increase the energy to the self-similar one, then the gravitational potential energy will dominate the energetics and the shock will likely stall and dissipate at a finite radius.

From Figure \ref{fig:ebofr}, when $n \lesssim 2.86$, there is a region within the shocked fluid where both the specific energy and Bernoulli parameter are positive, and as $n$ nears 2 this region encompasses more of the fluid and the energy itself grows to large values. Importantly, however, this does not imply that gas parcels are actually unbound from the black hole, and one can show that the \emph{entire fluid} will always fall back to the black hole at late enough times, independent of the adiabatic index of the gas. This finding implies that, in failed supernovae where this self-similar velocity scaling is maintained, the shock will not actually unbind material until it encounters a region of more rapidly declining density, which is typically only very near the surface of the star. It should then be possible to use the time-dependent energy from our self-similar solutions to predict the location at which the shock starts to unbind material from the black hole in a failed supernova, and hence the total amount of mass successfully ejected from the system. Once the shock starts to successfully unbind matter, the return of only marginally-bound material to the black hole will cause the fallback rate to transition to $\dot{M} \propto t^{-5/3}$.

\acknowledgements
We thank Rodrigo Fern\'andez for the results from the simulations of failed supernovae. We thank Chris Matzner, Chris McKee{}{, and the referee, David Weinberg,} for useful comments and suggestions. ERC acknowledges support from NASA through the Einstein Fellowship Program, grant PF6-170150. This work was supported in part by a Simons Investigator award from the Simons Foundation (EQ) and the Gordon and Betty Moore Foundation through Grant GBMF5076. 

\software{MESA \citep{paxton11, paxton13, paxton15, paxton18}, FLASH \citep{fryxell00}}

\appendix
\label{appendix:appendix}
{}{Our self-similar definitions for the velocity, density, and pressure (Equations \ref{vss} -- \ref{pss}) were written in terms of Eulerian $r$ and self-similar $r/r_{\rm sh}(t)$, which was motivated by the fact that the gravitational field is itself a function of $r$. We could have also, however, written our self-similar variables in terms of $t$ and $r/r_{\rm sh}(t)$, which would have less obviously satisfied the differential equations in a self-similar way, but would have been more in line with past investigations of self-similar shock propagation. For completeness, in terms of these variables the velocity, density, and pressure are}

\begin{equation}
v = v_{\rm sh}(t)\tilde{f}(\xi), \label{ssv1}
\end{equation}
\begin{equation}
\rho = \left(\frac{r_{\rm sh}(t)}{r_0}\right)^{-n}\tilde{g}(\xi),
\end{equation}
\begin{equation}
p = \left(\frac{r_{\rm sh}(t)}{r_0}\right)^{-n}V(t)^2\tilde{h}(\xi), \label{ssp1}
\end{equation}
{}{where the shock velocity and position are (as before)}

\begin{equation}
v_{\rm sh}(t) = V\sqrt{\frac{GM}{r_{\rm sh}(t)}}, \quad r_{\rm sh}(t) = \left(\frac{3}{2}V\sqrt{GM}t\right)^{2/3}.
\end{equation}
{}{Inserting Equations \eqref{ssv1} -- \eqref{ssp1} into the continuity, radial momentum, and entropy equations (Equations \ref{cont} -- \ref{gasen} in Section \ref{sec:equations}) then gives the following three ODEs for the modified, self-similar functions $\tilde{f}$, $\tilde{g}$, and $\tilde{h}$:}

\begin{equation}
-n\tilde{g}-\xi\tilde{g}'+\frac{1}{\xi^2}\frac{d}{d\xi}\left(\xi^2\tilde{g}\tilde{f}\right) = 0, \label{ss1p}
\end{equation}
\begin{equation}
-\frac{1}{2}\tilde{f}-\xi\tilde{f}+\tilde{f}\tilde{f}'+\frac{1}{\tilde{g}}\tilde{h}' = -\frac{1}{V^2\xi^2}, \label{ss2p}
\end{equation}
\begin{equation}
\left(\tilde{f}-\xi\right)\frac{d}{d\xi}\left(\frac{\tilde{h}}{\tilde{g}^{\gamma_2}}\right)+\left(-n-1+n\gamma_2\right)\frac{\tilde{h}}{\tilde{g}^{\gamma_2}} = 0. \label{ss3p}
\end{equation}
{}
{}{While writing the self-similar equations in this form does not yield any new information, it does make it more apparent that, if $n = 3$ and $\gamma_1 = \gamma_2 = 4/3$ -- appropriate to the case where the gas obeys a relativistic equation of state -- then Equations \eqref{ss1p} and \eqref{ss3p} can be integrated exactly to give}

\begin{equation}
\tilde{g} = \frac{C_1}{\xi^3-\xi^2\tilde{f}},
\end{equation}
\begin{equation}
\tilde{h} = C_2\tilde{g}^{4/3},
\end{equation}
{}{where $C_1 = c_{\rho}\left(1-c_v\right)$ and $C_2 = c_p/c_{\rho}^{4/3}$, and the $c$'s are given by Equations \eqref{cv} -- \eqref{cp} in Section \ref{sec:equations} with $\gamma_1 = \gamma_2 = 4/3$ and $n = 3$. Inserting these relations into Equation \eqref{ss2p} then yields the single differential equation for $\tilde{f}$:}

\begin{equation}
-\frac{1}{2}\tilde{f}-\xi\tilde{f}'+\tilde{f}\tilde{f}'+4C_2C_1^{1/3}\frac{d}{d\xi}\left[\xi\left(1-\frac{\tilde{f}}{\xi}\right)^{-1/3}\right] = -\frac{1}{V^2\xi^2}.
\end{equation}
{}{This expression can be rearranged to yield two algebraic relations for the location of the sonic point, the velocity $V$, and the value of the function $\tilde{f}$ at the sonic point that can be solved numerically.}

\bibliographystyle{aasjournal}

\begin{thebibliography}{}
\expandafter\ifx\csname natexlab\endcsname\relax\def\natexlab#1{#1}\fi
\providecommand{\url}[1]{\href{#1}{#1}}
\providecommand{\dodoi}[1]{doi:~\href{http://doi.org/#1}{\nolinkurl{#1}}}
\providecommand{\doeprint}[1]{\href{http://ascl.net/#1}{\nolinkurl{http://ascl.net/#1}}}
\providecommand{\doarXiv}[1]{\href{https://arxiv.org/abs/#1}{\nolinkurl{https://arxiv.org/abs/#1}}}

\bibitem[{{Chevalier}(1989)}]{chevalier89}
{Chevalier}, R.~A. 1989, \apj, 346, 847, \dodoi{10.1086/168066}

\bibitem[{{Coughlin}(2017)}]{coughlin17}
{Coughlin}, E.~R. 2017, \apj, 835, 40, \dodoi{10.3847/1538-4357/835/1/40}

\bibitem[{{Coughlin} {et~al.}(2018){Coughlin}, {Quataert}, {Fern{\'a}ndez}, \&
  {Kasen}}]{coughlin18}
{Coughlin}, E.~R., {Quataert}, E., {Fern{\'a}ndez}, R., \& {Kasen}, D. 2018,
  \mnras, \dodoi{10.1093/mnras/sty667}

\bibitem[{{Fern{\'a}ndez} {et~al.}(2018){Fern{\'a}ndez}, {Quataert},
  {Kashiyama}, \& {Coughlin}}]{fernandez18}
{Fern{\'a}ndez}, R., {Quataert}, E., {Kashiyama}, K., \& {Coughlin}, E.~R.
  2018, \mnras, 476, 2366, \dodoi{10.1093/mnras/sty306}

\bibitem[{{Fryxell} {et~al.}(2000){Fryxell}, {Olson}, {Ricker}, {Timmes},
  {Zingale}, {Lamb}, {MacNeice}, {Rosner}, {Truran}, \& {Tufo}}]{fryxell00}
{Fryxell}, B., {Olson}, K., {Ricker}, P., {et~al.} 2000, \apjs, 131, 273,
  \dodoi{10.1086/317361}

\bibitem[{{Hansen} {et~al.}(2004){Hansen}, {Kawaler}, \& {Trimble}}]{hansen04}
{Hansen}, C.~J., {Kawaler}, S.~D., \& {Trimble}, V. 2004, {Stellar interiors :
  physical principles, structure, and evolution}

\bibitem[{{Kazhdan} \& {Murzina}(1994)}]{kazhdan94}
{Kazhdan}, Y.~M., \& {Murzina}, M. 1994, \mnras, 270, 351,
  \dodoi{10.1093/mnras/270.2.351}

\bibitem[{{Koo} \& {McKee}(1990)}]{koo90}
{Koo}, B.-C., \& {McKee}, C.~F. 1990, \apj, 354, 513, \dodoi{10.1086/168712}

\bibitem[{{Lovegrove} \& {Woosley}(2013)}]{lovegrove13}
{Lovegrove}, E., \& {Woosley}, S.~E. 2013, \apj, 769, 109,
  \dodoi{10.1088/0004-637X/769/2/109}

\bibitem[{{Lovegrove} {et~al.}(2017){Lovegrove}, {Woosley}, \&
  {Zhang}}]{lovegrove17}
{Lovegrove}, E., {Woosley}, S.~E., \& {Zhang}, W. 2017, \apj, 845, 103,
  \dodoi{10.3847/1538-4357/aa7b7d}

\bibitem[{{Matzner} \& {McKee}(1999)}]{matzner99}
{Matzner}, C.~D., \& {McKee}, C.~F. 1999, \apj, 510, 379,
  \dodoi{10.1086/306571}

\bibitem[{{Nadezhin}(1980)}]{nadezhin80}
{Nadezhin}, D.~K. 1980, \apss, 69, 115, \dodoi{10.1007/BF00638971}

\bibitem[{{Ostriker} \& {McKee}(1988)}]{ostriker88}
{Ostriker}, J.~P., \& {McKee}, C.~F. 1988, Reviews of Modern Physics, 60, 1,
  \dodoi{10.1103/RevModPhys.60.1}

\bibitem[{{Paxton} {et~al.}(2011){Paxton}, {Bildsten}, {Dotter}, {Herwig},
  {Lesaffre}, \& {Timmes}}]{paxton11}
{Paxton}, B., {Bildsten}, L., {Dotter}, A., {et~al.} 2011, \apjs, 192, 3,
  \dodoi{10.1088/0067-0049/192/1/3}

\bibitem[{{Paxton} {et~al.}(2013){Paxton}, {Cantiello}, {Arras}, {Bildsten},
  {Brown}, {Dotter}, {Mankovich}, {Montgomery}, {Stello}, {Timmes}, \&
  {Townsend}}]{paxton13}
{Paxton}, B., {Cantiello}, M., {Arras}, P., {et~al.} 2013, \apjs, 208, 4,
  \dodoi{10.1088/0067-0049/208/1/4}

\bibitem[{{Paxton} {et~al.}(2015){Paxton}, {Marchant}, {Schwab}, {Bauer},
  {Bildsten}, {Cantiello}, {Dessart}, {Farmer}, {Hu}, {Langer}, {Townsend},
  {Townsley}, \& {Timmes}}]{paxton15}
{Paxton}, B., {Marchant}, P., {Schwab}, J., {et~al.} 2015, \apjs, 220, 15,
  \dodoi{10.1088/0067-0049/220/1/15}

\bibitem[{{Paxton} {et~al.}(2018){Paxton}, {Schwab}, {Bauer}, {Bildsten},
  {Blinnikov}, {Duffell}, {Farmer}, {Goldberg}, {Marchant}, {Sorokina},
  {Thoul}, {Townsend}, \& {Timmes}}]{paxton18}
{Paxton}, B., {Schwab}, J., {Bauer}, E.~B., {et~al.} 2018, \apjs, 234, 34,
  \dodoi{10.3847/1538-4365/aaa5a8}

\bibitem[{{Piro}(2013)}]{piro13}
{Piro}, A.~L. 2013, \apjl, 768, L14, \dodoi{10.1088/2041-8205/768/1/L14}

\bibitem[{{Ro} \& {Matzner}(2013)}]{ro13}
{Ro}, S., \& {Matzner}, C.~D. 2013, \apj, 773, 79,
  \dodoi{10.1088/0004-637X/773/1/79}

\bibitem[{{Sakurai}(1960)}]{sakurai60}
{Sakurai}, A. 1960, CPAM, 13, 353

\bibitem[{{Sedov}(1959)}]{sedov59}
{Sedov}, L.~I. 1959, {Similarity and Dimensional Methods in Mechanics}

\bibitem[{{Taylor}(1950)}]{taylor50}
{Taylor}, G. 1950, Proceedings of the Royal Society of London Series A, 201,
  159, \dodoi{10.1098/rspa.1950.0049}

\bibitem[{{Waxman} \& {Shvarts}(1993)}]{waxman93}
{Waxman}, E., \& {Shvarts}, D. 1993, Physics of Fluids A, 5, 1035,
  \dodoi{10.1063/1.858668}

\end{thebibliography}

\end{document}